\documentclass[preprint2,twocolappendix,x11names]{aastex631_arXiv}

\shorttitle{BHB Stars with LAMOST}
\shortauthors{Wang et al.}
\usepackage{amsmath}
\usepackage{longtable}
\usepackage{booktabs}
\usepackage{multirow}
\usepackage{comment}
\usepackage{CJKfntef}

\begin{document}

\title{Identifying and Determining Atmospheric Parameters of BHB Stars Based on LAMOST DR11}

\correspondingauthor{Wen-Yuan Cui}
\email{cuiwenyuan@hebtu.edu.cn}

\author[0000-0003-2536-3142]{Xiao-Long Wang}
\altaffiliation{Physics Postdoctoral Research Station at Hebei Normal University.}
\affiliation{Department of Physics, Hebei Normal University, Shijiazhuang 050024, People's Republic of China; xlwang@hebtu.edu.cn}
\affiliation{Guo Shoujing Institute for Astronomy, Hebei Normal University, Shijiazhuang 050024, People's Republic of China}
\affiliation{Hebei Advanced Thin Films Laboratory, Shijiazhuang 050024, People's Republic of China}

\author[0000-0003-1359-9908]{Wen-Yuan Cui}
\affiliation{Department of Physics, Hebei Normal University, Shijiazhuang 050024, People's Republic of China; xlwang@hebtu.edu.cn}
\affiliation{Guo Shoujing Institute for Astronomy, Hebei Normal University, Shijiazhuang 050024, People's Republic of China}

\author{Jie Ju}
\affiliation{Hebei University of Science and Technology, Shijiazhuang 050018, People's Republic of China}

\author[0000-0003-1828-5318]{Guo-Zhen Hu}
\altaffiliation{Physics Postdoctoral Research Station at Hebei Normal University.}
\affiliation{Department of Physics, Hebei Normal University, Shijiazhuang 050024, People's Republic of China; xlwang@hebtu.edu.cn}
\affiliation{Guo Shoujing Institute for Astronomy, Hebei Normal University, Shijiazhuang 050024, People's Republic of China}
\affiliation{Hebei Key Laboratory of Photophysics Research and Application, Shijiazhuang 050024, People's Republic of China}

\author[0000-0001-8060-1321]{Min Fang}
\affiliation{Purple Mountain Observatory, Chinese Academy of Sciences, Nanjing 210023, People's Republic of China}
\affiliation{School of Astronomy and Space Science, University of Science and Technology of China, Hefei 230026, People's Republic of China}

\author[0000-0002-4799-0780]{Shuai Zhang}
\affiliation{Department of Physics, Hebei Normal University, Shijiazhuang 050024, People's Republic of China; xlwang@hebtu.edu.cn}
\affiliation{Guo Shoujing Institute for Astronomy, Hebei Normal University, Shijiazhuang 050024, People's Republic of China}

\author[0000-0002-4828-0326]{Jia-Ming Liu}
\affiliation{Department of Physics, Hebei Normal University, Shijiazhuang 050024, People's Republic of China; xlwang@hebtu.edu.cn}
\affiliation{Guo Shoujing Institute for Astronomy, Hebei Normal University, Shijiazhuang 050024, People's Republic of China}

\begin{abstract}
Large catalogs of blue horizontal-branch (BHB) stars are essential for studying substructures and kinematics of the Galactic halo. And accurate determination of atmospheric parameters for BHB stars provides insight into stellar evolution. In this work, we perform a systematic search for BHB stars based on LAMOST DR11, and identify 13\,988 BHB spectra, corresponding to 10\,236 unique BHB stars. We estimate an identification rate of $\sim$80\%\textendash90\%, and a contamination rate of $\lesssim$10\% for our sample. Atmospheric parameters for these BHB stars are estimated via the data-driven method named the Stellar LAbel Machine (\texttt{SLAM}). We demonstrate the necessity of including color indices in the spectral labeling to effectively break the degeneracy between effective temperature and surface gravity. We note a bump in the distribution of [Fe/H], and most of these metal-rich BHB stars belong to the disk population. We also provide a list of 4282 blue straggler (BS) stars with determined atmospheric parameters.
\end{abstract}

\keywords{horizontal branch stars --- atmospheric parameters --- Catalogs --- Spectroscopy --- LAMOST}

\section{Introduction}\label{sec:intro}

Blue horizontal-branch (BHB) stars are horizontal-branch (HB) stars falling to the blue of the RR Lyrae instability strip on the Hertzsprung-Russell (HR) diagram, and they were first unambiguously detected and defined  in Galactic globular clusters M~92 and M~3 through photometric studies \citep{Arp1952AJ.....57....4A,Sandage1953AJ.....58...61S}. A low mass star with initial mass of $\lesssim1\,M_{\odot}$ will evolve off the main-sequence (MS) once core hydrogen burning ceases, and the star evolves onto the red-giant branch (RGB) burning hydrogen in its shell and displaying appreciable stellar wind. The inert helium core becomes electron-degenerate during contraction, and the temperature in the core increases. Once the temperature is high enough, the core helium burning begins and the helium flash occurs, increasing the temperature in the core further and rendering the core non-degenerate. Then the star moves to the blue horizontal-branch, becoming a so-called BHB star \citep{Rood1973ApJ...184..815R,Catelan2009Ap&SS.320..261C}. BHB stars are old, metal-poor giant stars with masses of $<0.8\,M_{\odot}$ \citep{Heber1997ESASP.402..461H,Catelan2009Ap&SS.320..261C},  and burning helium in their cores and hydrogen in theirs shells \citep{Ruhland2011ApJ...731..119R,Paunzen2019A&A...622A..77P}.

Though depending slightly on both metallicity and temperature \citep{Wilhelm1999AJ....117.2308W,Sirko2004AJ....127..899S}, BHB stars have nearly constant absolute magnitudes at given colors, making them a particular useful standard candle in determining distances \citep[e.g.,][]{Clewley2006MNRAS.368..310C,Xue2008ApJ...684.1143X}. In addition, BHB stars have large luminosities, allowing them to be detected at large distances. These properties make BHB stars good tracers for studying substructures and kinematics in the Galactic Halo \citep[e.g.,][]{Xue2008ApJ...684.1143X,Xue2011ApJ...738...79X,Deason2011MNRAS.416.2903D,Starkenburg2019MNRAS.490.5757S,Utkin2020MNRAS.499.1058U}. For example, \citet{Xue2008ApJ...684.1143X} constrained the mass distribution of our Milky Way's dark matter halo out to a galactocentric distance of $\sim$60\,kpc by analyzing the kinematics of $\sim$2400 BHB stars, and \citet{Xue2011ApJ...738...79X} found that the outer halo is more substructured than the inner halo based on a sample of over 4000 BHB stars.

In addition to being more luminous than most giant-branch stars and having small scatters in intrinsic luminosities, BHB stars display spectral features rendering them identifiable. BHB stars are among objects displaying the strongest Balmer jump and display deep Balmer lines, due to their effective temperatures \citep[e.g.,][$T_{\rm eff}\sim7000\textendash12000\,{\rm K}$]{Santucci2015ApJ...801..116S,Ju2024ApJS..270...11J,Ju2025ApJS..276...12J} making the level population of hydrogen atoms peaks at level $n=2$, and their low surface gravities \citep[e.g.,][$\log g\sim3\textendash4$]{Santucci2015ApJ...801..116S,Ju2024ApJS..270...11J,Ju2025ApJS..276...12J}. Their low surface gravities also render their Balmer lines narrower than main-sequence A-type stars.

Observationally, BHB stars are identified either photometrically or spectroscopically utilizing these spectral features. The large Balmer jump in BHB stars makes them appear redder in $u-g$ colors than objects such as white dwarfs and low-redshift quasars. In addition, BHB stars are bluer in $g-r$ colors than most halo stars, since the main-sequence turn-off for halo stars lies at spectral types F and G. These color properties make photometric selection of halo BHB stars straightforward, and color cuts have been used as initial selection of BHB candidates in various studies, followed by further spectroscopic confirmation \citep[e.g.,][]{Sirko2004AJ....127..899S,Xue2008ApJ...684.1143X,Bell2010AJ....140.1850B,Xue2011ApJ...738...79X}. \citet{Lenz1998ApJS..119..121L} found that the $i-z$ color space can also be used to separate A-type stars with different surface gravities, though less effective than the $u-g$ color space. \citet{Vickers2012AJ....143...86V} demonstrated that the splitting in this color space is caused by the Paschen features residing in the $z$-band, and reported a purity of $\sim$77\% for selecting samples of BHB stars using $i-z$ colors. Recently, \citet{Hu2025ApJS..278...62H} noted that the $u-v$ color in the SkyMapper photometric system \citep{Bessell2011PASP..123..789B} is a reliable proxy for surface gravity, and constructed a catalog of $\sim$50\,000 BHB stars based on synthetic colors convolved from the ``corrected" Gaia XP spectra \citep{Huang2024ApJS..271...13H}. 

While photometric studies are capable of picking up BHB stars at great distances (e.g., the Magellanic Clouds at distances of $\sim$50\textendash60\,kpc, in \citealt{Belokurov2016MNRAS.456..602B}, or the Galactic halo out to hundreds of kpc, in \citealt{Nie2015ApJ...810..153N,Thomas2018MNRAS.481.5223T,Fukushima2018PASJ...70...69F, Fukushima2019PASJ...71...72F}), spectroscopic follow-up is essential for obtaining high purity BHB star catalogs \citep[e.g.,][]{Sirko2004AJ....127..899S,Clewley2004MNRAS.352..285C,Xue2008ApJ...684.1143X,Xue2011ApJ...738...79X,Deason2011MNRAS.416.2903D}. Generally, spectroscopic studies yield BHB star catalogs with contamination rates well below 10\% \citep[e.g.,][]{Xue2008ApJ...684.1143X}, whereas photometrically selected samples suffer $\sim$30\% contamination (e.g., \citealt{Bell2010AJ....140.1850B}; although \citealt{Starkenburg2019MNRAS.490.5757S,Hu2025ApJS..278...62H} reported completeness and purity both exceeding 90\% in their samples, utilizing specialized photometric filters). Recently, \citet{Ju2024ApJS..270...11J} compiled a catalog of over 5000 BHB stars based solely on analyzing Balmer line profiles in low resolution spectra from LAMOST DR5, and \citet{Bystrom2025MNRAS.542..560B} constructed a sample of $\sim$10\,000 BHB stars using atmospheric parameters from the Dark Energy Spectroscopic Instrument (DESI) survey \citep{DESI-Collaboration2022AJ....164..207D}. In addition, \citet{Vickers2021ApJ...912...32V} constructed a sample of more than 13\,000 BHB stars by applying a machine-learning algorithm to the LAMOST spectra, with BHB stars from \citet{Brown2008AJ....135..564B} and \citet{Xue2008ApJ...684.1143X} as training set.

Large catalogs of BHB stars with high reliability are essential for studying the structures and kinematics of the Galaxy \citep[e.g.,][]{Xue2008ApJ...684.1143X,Xue2011ApJ...738...79X,Utkin2020MNRAS.499.1058U}, and serving as fundamental input to machine learning methods \citep[e.g.,][]{Vickers2021ApJ...912...32V,Wei2023PASP..135h4501W,Zhang2025AJ....170..158Z}. In this paper, we aim to identify more BHB stars spectroscopically, and determine their atmospheric parameters based on the new data release from the LAMOST survey. The paper is organized as follows. In Section~\ref{sec:data}, we describe the data sets used in this paper and the preprocessing of the LAMOST spectra. The identification method is described in Section~\ref{sec:BHBinden}, and in Section~\ref{sec:BHBatmos}, we determine the atmospheric parameters of the BHB stars. In Section~\ref{sec:dis}, we discuss some properties of the BHB sample. Finally, a summary section is given in Section~\ref{sec:sum}.

\section{Data Sets and Spectral Processing}\label{sec:data}
\subsection{Optical Spectroscopic Data from the LAMOST Survey}\label{sec:data_lamost}

The main data set used in this work is the spectroscopic data from the LAMOST\footnote{The Large Sky Area Multi-Object Fiber Spectroscopic Telescope, also called the Guoshoujing Telescope.} survey \citep{Cui2012RAA....12.1197C}. LAMOST is located at Xinglong Observatory Station in Hebei province, China. The telescope has an effective aperture of $\sim$4\,\rm m and a field of view of $5^{\circ}$ in diameter. There are 16 spectrographs and 4000 fibers equipped on the telescope. The spectrographs have a resolution of $R\approx1800$ and covers a wavelength range of 3700\textendash9100\,$\rm\AA$ \citep{Cui2012RAA....12.1197C,Zhao2012RAA....12..723Z,Liu2015RAA....15.1089L,Luo2015RAA....15.1095L}. Since 2018, the spectrographs have been upgraded to support median resolution observations as well, with $R\approx7500$, and wavelength coverages of 4950\textendash5350\,$\rm\AA$ in the blue arm and 6300\textendash6800\,$\rm\AA$ in the red arm \citep{Liu2020arXiv200507210L}.

By the time we start this work, the latest release is LAMOST DR11\footnote{\url{https://www.lamost.org/dr11/}}. LAMOST DR11 contains more than 11.9 million spectra, about 11.6 million of which are stellar spectra. In this work, we consider only spectra have signal-to-noise ratios in the $g$-band (SNRg) larger than 10. In addition, stars close to the Galactic plane suffer larger extinction, making assessment of their properties less reliable, so we exclude stars with $|b|<20^{\circ}$. Finally, there are 4\,945\,922 spectra with $\rm SNRg\ge10$ and $|b|\ge20^{\circ}$ in LAMOST DR11, and these spectra constitute the ``parent sample" analyzed in this work.

\subsection{Additional Ancillary Data}\label{sec:data_other}

In this work, we also include the astrometric measurements and photometry from the 3rd data release of the Gaia satellite \citep[Gaia DR3;][]{Gaia-Collaboration2016A&A...595A...1G,Gaia-Collaboration2023A&A...674A...1G}, and $JH$ photometry from the Two Micron All Sky Survey \citep[2MASS;][]{Skrutskie_2MASS_IPAC,Skrutskie2006AJ....131.1163S}. Photometry with magnitude errors $>$0.1\,mag are omitted. The photometric data serve as supplement information to break the degeneracy between effective temperature and surface gravity when determine atmospheric parameters in Section~\ref{sec:BHBatmos}.

\subsection{Preprocessing of the LAMOST Spectra}\label{sec:data_spec_red}

For our analysis, we firstly shift each spectrum in the parent sample to the rest frame using the radial velocity (RV) values in the LAMOST catalog.

In addition, it has been noted that bad pixels are quite common in spectroscopic surveys due to sky subtraction, cosmic rays, and problems occurring during data reduction \citep{Zhang2020ApJS..246....9Z}. These bad pixels contain no meaningful information about the star and may bias our interpretation of the target's intrinsic properties, so they should be removed from further analysis. Specifically, the LAMOST spectra provide \texttt{ORMASK} for every pixel in the spectrum, with zero values indicating no problems in any single exposure, and non-zero integers otherwise (see \url{https://www.lamost.org/dr11/v1.1/doc/lr-data-production-description} for more description). In our analysis, we exclude these bad pixels with non-zero \texttt{ORMASK} values. We note that there remain pixels with non-positive fluxes or zero ``inverse variance'' (one over sigma-squared, and sigma is the uncertainty of the corresponding pixel) in some spectra, and these pixels are removed as well. There are also spike-like pixels and extreme dippers in the spectrum, due to imperfect cosmic ray removal and sky subtraction. We use the \texttt{find\_peaks}\footnote{\url{https://docs.scipy.org/doc/scipy/reference/generated/scipy.signal.find\_peaks.html}} function from the python package \texttt{scipy} \citep{Virtanen2020NatMe..17..261V} to search these extremely narrow features and exclude them from our analysis.

In the following, all measurements and analyses are performed on these preprocessed spectra.

\section{Identification of BHB Stars}\label{sec:BHBinden}

\subsection{Selection of A-type Stars}\label{sec:Atype}

\begin{figure}[!t]
    \centering
    \includegraphics[width=\columnwidth]{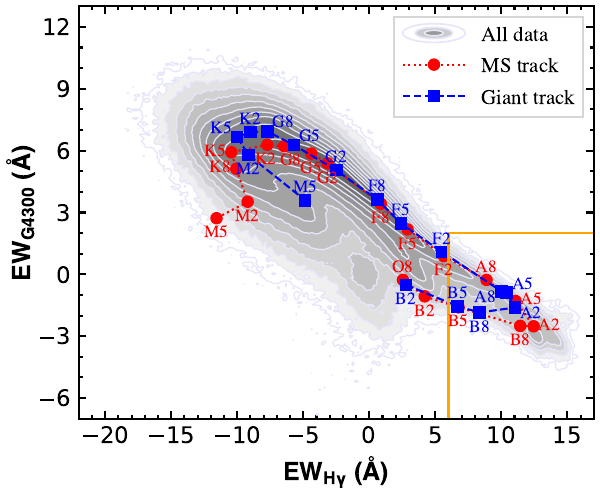}
    \caption{The background image and contours display the distribution of all spectra with $\rm SNRg\ge10$ and $|b|\ge20^{\circ}$ in the $\rm EW_{G4300}$ vs. $\rm EW_{H\gamma}$ plane. The red circles connected with red dotted curve represent the main-sequence track, and the giant track is shown as blue squares connected with blue dashed curve \citep[adopted from][]{Liu2015RAA....15.1137L}. The orange solid lines denote cuts on $\rm EW_{G4300}$ and $\rm EW_{H\gamma}$ for the selection of A-type stars.}
    \label{fig:EWG4300_vs_EWHgamma}
\end{figure}

Observationally, there have been a variety of sub-categories of BHB stars, including A-type horizontal-branch (HBA) stars that are cooler than 12\,000\,K, B-type horizontal-branch (HBB) stars with effective temperatures in the range 12\,000\textendash20\,000\,K, and extreme horizontal-branch (EHB) stars that are hotter than 20\,000\,K \citep{Newell1970ApJ...159..443N,Heber1986A&A...162..171H,Mohler2004IAUS..224..395M,Catelan2009Ap&SS.320..261C,Heber2009ARA&A..47..211H}. In this work, we focus mainly on the sub-category HBA stars, and in the remainder of the text, the term ``BHB'' always refers to the HBA sub-category. Their spectra are characterized by strong hydrogen lines and weak or no molecular bands, which could help separating them from cool or hot contaminants. \citet{Liu2015RAA....15.1137L} found that the stars with all spectral types from O to M type can be well separated and ordered in the $\rm EW_{G4300}$ versus  $\rm EW_{H\gamma}$ plane, where $\rm EW_{H\gamma}$ is the equivalent width of the H$\gamma$ line, and $\rm EW_{G43000}$ denotes the line index of the molecular band around 4300\,\AA. We use the \texttt{measure\_line\_index}\footnote{\url{https://laspec.readthedocs.io/en/latest/api.html\#laspec.line\_index.measure\_line\_index}} function in the PYTHON package \texttt{laspec} \citep{Zhang2020ApJS..246....9Z,Zhang2021ApJS..256...14Z} and follow the line definitions in \citet{Liu2015RAA....15.1137L} to measure $\rm EW_{H\gamma}$ and $\rm EW_{G4300}$. In Figure~\ref{fig:EWG4300_vs_EWHgamma}, we display all the spectra in the parent sample (see Section~\ref{sec:data_lamost}), as well as the empirical stellar loci of main-sequence and giant stars, in the $\rm EW_{G4300}$ versus $\rm EW_{H\gamma}$ plane. From the plot, A-type stars occupy the lower-right portion of the diagram, i.e., with smaller values of $\rm EW_{G4300}$, and larger values of $\rm EW_{H\gamma}$. We empirically define the cuts (shown as orange solid lines in Figure~\ref{fig:EWG4300_vs_EWHgamma}) to cover almost all of the A-type stars as follows:
\begin{equation}
\rm EW_{H\gamma}\ge6\,\AA,\ EW_{G4300}\leq2\,\AA.
\end{equation}

\noindent With these cuts, we obtain 91\,562 A-type spectra as BHB candidates, in which the main contaminants are blue straggler (BS) stars, cool A-type MS stars, and early F-type stars. Indeed, we note that more than half of the candidates are classified as late A- or early F-types by the LAMOST pipeline \citep{Luo2012RAA....12.1243L,Luo2015RAA....15.1095L}. These contaminants will be removed in Section~\ref{sec:Balmer}, using the $D_{0.2}$ method and the scale width-shape method.

\subsection{Balmer Line Analyses}\label{sec:Balmer}

\begin{figure*}[!t]
    \centering
    \includegraphics[width=\textwidth]{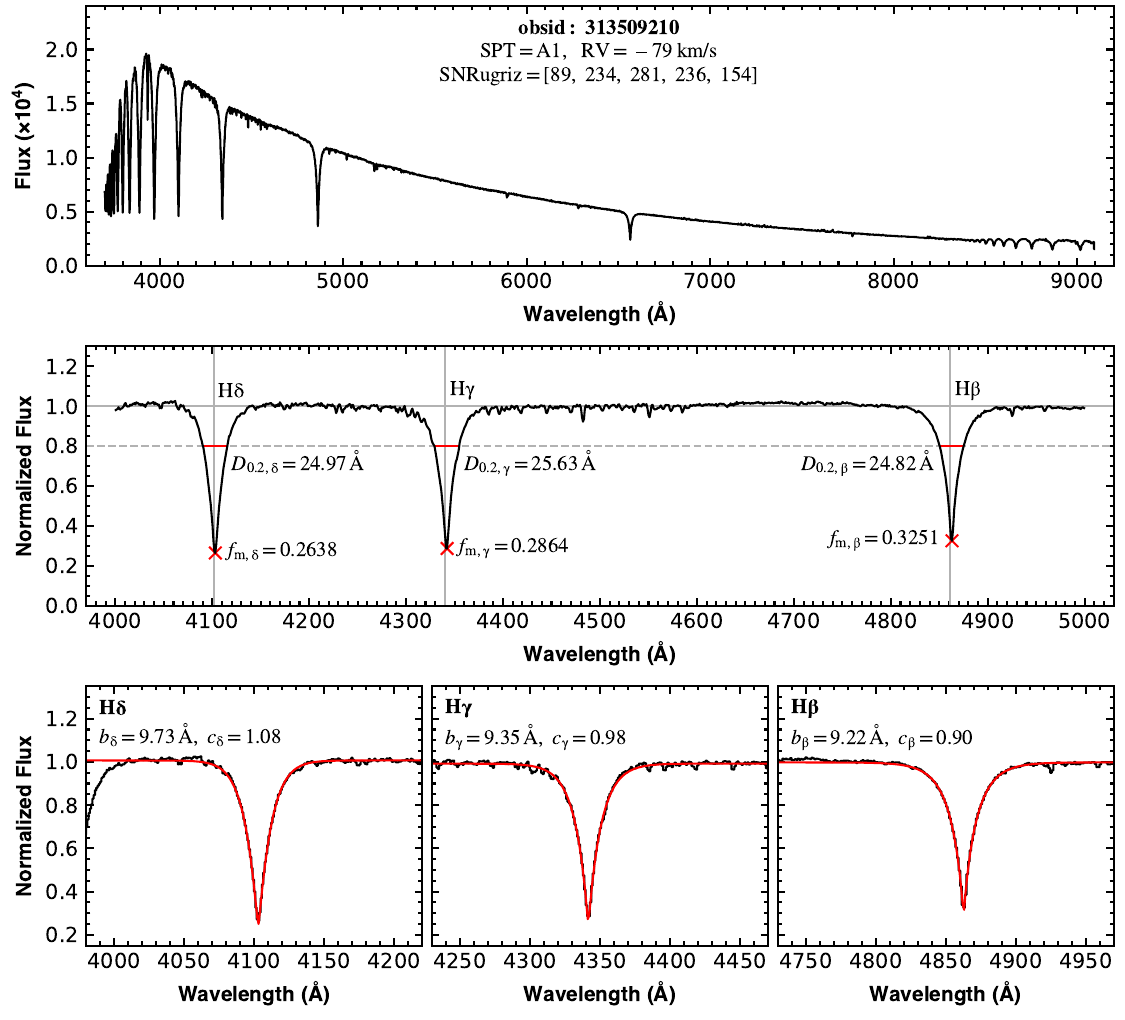}
    \caption{Top panel: An example LAMOST spectrum of a typical BHB star. The spectrum has been corrected to the rest frame, with bad pixels and outlier measurements removed (see Section~\ref{sec:data_spec_red} for details). Some basic information for the spectrum provided in the LAMOST catalog, including the unique spectrum ID number (\texttt{obsid}), the spectral type, the radial velocity and the signal-to-noise ratios in the $ugriz$-bands, are labeled. Middle panel: Normalized spectrum zoomed into the spectral range covering the H$\beta$, H$\gamma$ and H$\delta$ lines (marked with gray vertical lines). The gray dashed horizontal line shows the level at where we measure $D_{0.2}$, and the red line segments denote the wavelength spanning of $D_{0.2}$ for the three Balmer lines. The red crosses mark the level at where the $f_{\rm m}$ values are measured. The measured $D_{0.2}$ and $f_{\rm m}$ are labeled for the three Balmer lines. Bottom panels: Fitting S{\'e}rsic profile to the three Balmer lines. The observed profile is shown as black and the best fitted S{\'e}rsic profile is shown as red. The best fitted scale width ($b$) and scale shape ($c$) are labeled.}
    \label{fig:BHBspec_example}
\end{figure*}

As mentioned in Section~\ref{sec:Atype}, we focus mainly on the HBA sub-category of BHB stars herein, that have effective temperatures in the range 7000\textendash12\,000\,K. In this temperature range, BHB stars have lower surface gravities than BS stars. The Balmer line profiles of stars in this temperature range are sensitive to both surface gravity and effective temperature. Thus their analyses serve as powerful tools of selecting BHB stars with confidence. In Figure~\ref{fig:BHBspec_example}, we present a typical BHB spectrum, displaying deep Balmer lines. In this section, we apply two methods to identify BHB stars: the $D_{0.2}$ method \citep[e.g.,][]{Pier1983ApJS...53..791P,Sommer-Larsen1986MNRAS.219..537S,Arnold1992MNRAS.257..225A,Flynn1994MNRAS.267...77F,Kinman1994AJ....108.1722K,Wilhelm1999AJ....117.2308W,Sirko2004AJ....127..899S} and the scale width-shape method \citep{Clewley2002MNRAS.337...87C}. We focus on analyses of the H$\beta$, H$\gamma$ and H$\delta$ lines, because H$\alpha$ line have lower continuum level and higher order Balmer lines suffer severe line crowding, making the fit of the continuum level difficult.

To analyze the Balmer lines, the spectrum must be normalized to the continuum level. In this section, we focus on the wavelength range from 4000 to 5000\,$\rm\AA$ covering the three Balmer lines. Similar as in \citet{Sirko2004AJ....127..899S}, a sixth-order Legendre polynomial is fitted to this spectral range with 120\,$\rm\AA$ masks centered on the three Balmer lines excluded, to obtain the continuum level. Then the spectrum is normalized by dividing the best-fit continuum. An example normalized spectrum is shown in the middle panel of Figure~\ref{fig:BHBspec_example}. In Sections~\ref{sec:D20_vs_fm} and \ref{sec:b_vs_c}, the measurements of the Balmer line properties are performed on the normalized spectra.

\subsubsection{The $D_{0.2}$ Method}\label{sec:D20_vs_fm}

Since BHB stars are giant stars with low surface gravity, their Balmer lines are narrower than BS stars. The standard way of discriminating BHB stars from BS stars is measuring the line width of the Balmer lines. There have been different methods of measuring the line width (e.g., the $D_{0.2}$ method first propsed by \citealt{Pier1983ApJS...53..791P}, and the $D_{0.15}$ method in \citealt{Clewley2002MNRAS.337...87C}). In this work, we adopt the widely used $D_{0.2}$ method to select BHB stars \citep[e.g.,][]{Sirko2004AJ....127..899S,Xue2008ApJ...684.1143X,Xue2011ApJ...738...79X,Ju2024ApJS..270...11J}. The $D_{0.2}$ method measures the value of $D_{0.2}$, i.e., the width of the Balmer line at the level 20\% below the continuum level (as illustrated with the red line segments in the middle panel of Figure~\ref{fig:BHBspec_example}).

\begin{figure*}[!t]
    \centering
    \includegraphics[width=0.9\textwidth]{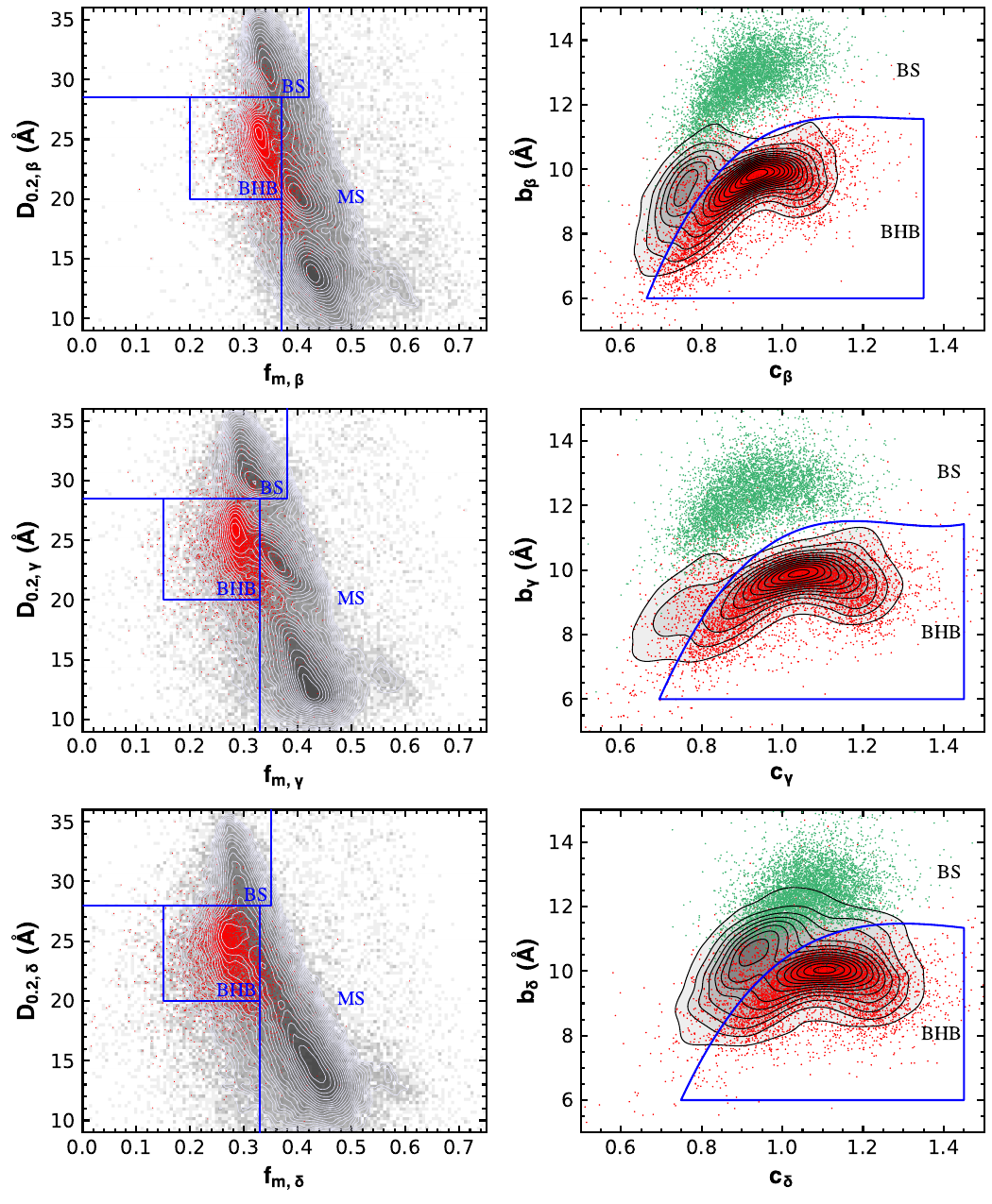}
    \caption{Top row: Identifying BHB candidates based on the $D_{0.2}$ method (top left) and the scale width-shape method (top right) using the H$\beta$ line. In both panels, the red dots represent BHB stars collected from various studies \citep[i.e.,][]{Xue2008ApJ...684.1143X,Xue2011ApJ...738...79X,Ju2024ApJS..270...11J}, and the blue lines show cuts on the parameters for selecting BHB and BS stars. In the top left panel, the background image with overlaid contours displays the distribution of A-type spectra identified in Section~\ref{sec:Atype}, and in the top right panel, it shows the distribution of BHB candidates identified via the $D_{0.2,\beta}$ method (i.e., points fall in the blue box in the top left panel). Middle and bottom rows are the same as the top row, but for the H$\gamma$ and H$\delta$ lines, respectively. The green dots in the right column represent BS candidates identified via Equation~\ref{eq:D20_for_BS}.}
    \label{fig:BalmerLines_for_BHBiden}
\end{figure*}

Traditionally, the $D_{0.2}$ method applies only to stars with A-type spectra \citep[e.g.,][]{Pier1983ApJS...53..791P,Sommer-Larsen1986MNRAS.219..537S}. However, as mentioned in Section~\ref{sec:Atype}, there are contaminants of early F and late A-type stars in our sample. These cool stars have weaker Balmer lines than BHB stars, and can be eliminated by measuring the line depths. Following \citet{Sirko2004AJ....127..899S}, we define $f_{\rm m}$ as the minimum flux of the line in the normalized spectrum (as marked with red crosses in the middle panel of Figure~\ref{fig:BHBspec_example}), and thus stars with higher values of $f_{\rm m}$ are cooler.

The measured values of $D_{0.2}$ and $f_{\rm m}$ are displayed in the left panels of Figure~\ref{fig:BalmerLines_for_BHBiden} for the three Balmer lines. As shown, there are clear density peaks around $(f_{{\rm m},\beta},\ D_{0.2,\beta})=(0.32,\ 25.6\,{\rm\AA})$, $(f_{{\rm m},\gamma},\ D_{0.2,\gamma})=(0.28,\ 26.0\,{\rm\AA})$, and $(f_{{\rm m},\delta},\ D_{0.2,\delta})=(0.27,\ 25.5\,{\rm\AA})$, and these peaks almost coincide with the distribution of the known BHB stars \citep[compiled from][]{Xue2008ApJ...684.1143X,Xue2011ApJ...738...79X,Ju2024ApJS..270...11J}, though the peak is less prominent using the H$\delta$ line. There are also density peaks at $D_{0.2,\beta}\gtrsim28.5\,{\rm\AA}$, $D_{0.2,\gamma}\gtrsim28.5\,{\rm\AA}$, and $D_{0.2,\delta}\gtrsim28.0\,{\rm\AA}$, and these peaks are mainly BS stars \citep{Sirko2004AJ....127..899S,Xue2008ApJ...684.1143X}, again the peak is less prominent using the H$\delta$ line. In the panel using the H$\beta$ line, there are two additional density peaks at $(f_{{\rm m},\beta},\ D_{0.2,\beta})=(0.40,\ 20.3\,{\rm\AA})$ and $(f_{{\rm m},\beta},\ D_{0.2,\beta})=(0.43,\ 13.7\,{\rm\AA})$. Checking the LAMOST catalog, these two peaks are dominated by A7/A6, and F0 types, respectively. There are similar peaks at $(f_{{\rm m},\gamma},\ D_{0.2,\gamma})=(0.36,\ 23.2\,{\rm\AA})$ and $(f_{{\rm m},\gamma},\ D_{0.2,\gamma})=(0.42,\ 12.0\,{\rm\AA})$, dominated by late A and early F-types, respectively, using the H$\gamma$ line. In the panel using the H$\delta$ line, the peak at $(f_{{\rm m},\delta},\ D_{0.2,\delta})=(0.43,\ 14.0\,{\rm\AA})$ is dominated by late A and F-types.

Based on the distribution in the $D_{0.2}$ versus $f_{\rm m}$ planes, and using the known BHB stars as references, we flag a spectrum as a BHB candidate if it meets any of the following criteria (shown as blue boxes in the left panels of Figure~\ref{fig:BHBspec_example}):
{\small\begin{equation}
\begin{cases}
20.0\,{\rm\AA}\leq D_{0.2,\beta}\leq28.5\,{\rm\AA},\ \text{and}\ 0.20\leq f_{{\rm m},\beta}\leq0.37;\\
20.0\,{\rm\AA}\leq D_{0.2,\gamma}\leq28.5\,{\rm\AA},\ \text{and}\ 0.15\leq f_{{\rm m},\beta}\leq0.33;\\
20.0\,{\rm\AA}\leq D_{0.2,\delta}\leq28.0\,{\rm\AA},\ \text{and}\ 0.15\leq f_{{\rm m},\gamma}\leq0.33.
\end{cases}
\end{equation}}

\noindent After applying these cuts, we obtain 15\,197, 13\,713, and 19\,470 BHB candidates using the H$\beta$, H$\gamma$, and H$\delta$ lines, respectively. With these cuts, most of the contaminants of BS stars and cool MS stars are removed.

As a by-product, we flag spectra meet all of the following criteria as BS candidates:
\begin{equation}
\begin{cases}
28.5\,{\rm\AA}\leq D_{0.2,\beta}\leq40.0\,{\rm\AA},\ \text{and}\ f_{{\rm m},\beta}\leq0.42;\\
28.5\,{\rm\AA}\leq D_{0.2,\gamma}\leq40.0\,{\rm\AA},\ \text{and}\ f_{{\rm m},\beta}\leq0.38;\\
28.0\,{\rm\AA}\leq D_{0.2,\delta}\leq40.0\,{\rm\AA},\ \text{and}\ f_{{\rm m},\gamma}\leq0.35.
\end{cases}\label{eq:D20_for_BS}
\end{equation}
We apply upper limits of $40\,\rm\AA$ for $D_{0.2,\beta}$, $D_{0.2,\gamma}$, and $D_{0.2,\delta}$ to remove contaminants of white dwarfs\footnote{We note that \citet{Lee2008AJ....136.2022L} flagged stars with $D_{0.2,\delta}>35\,\rm\AA$ as white dwarfs, and \citet{Guo2015MNRAS.454.2787G} used the criteria of $D_{0.2,\beta}>30\,\rm\AA$, $D_{0.2,\gamma}>50\,\rm\AA$, and $D_{0.2,\delta}>50\;\rm\AA$ to select white dwarfs. For simplicity, we adopt a unique upper limit of $40\,\rm\AA$ to remove white dwarfs.}. With these cuts, we obtain 6148 candidate BS spectra (green dots in the right panels of Figure~\ref{fig:BalmerLines_for_BHBiden}), and these candidates serve as references to further refine the selection of BHB stars via the scale width-shape method (see Section~\ref{sec:b_vs_c}).

\subsubsection{The Scale Width-Shape Method}\label{sec:b_vs_c}

The scale width-shape method \citep{Clewley2002MNRAS.337...87C} is based on fitting S{\'e}rsic profile \citep{Sersic1968adga.book.....S} to the Balmer lines,
\begin{equation}
y=1.0-a\exp\left[-\left(\dfrac{|\lambda-\lambda_{0}|}{b}\right)^{c}\right],
\end{equation}
where $y$ is the normalized flux, $\lambda$ is the wavelength, and $\lambda_{0}$ is the line center. The free parameters are $a$, $b$ and $c$, and $b$ and $c$ are called the scale width and scale shape, respectively. To account for uncertainties arising from radial velocity corrections and imperfect normalization of the spectra, we fit the normalized spectra to the S{\'e}rsic profile with five free parameters: $a$, $b$, $c$, $\lambda_{0}$, and $n$ following \citet{Xue2008ApJ...684.1143X},
\begin{equation}
y=n-a\exp\left[-\left(\dfrac{|\lambda-\lambda_{0}|}{b}\right)^{c}\right].
\end{equation}

\citet{Clewley2002MNRAS.337...87C} demonstrated that the scale shape $c$ provides measurements of the temperatures for A-type stars, with cooler stars having smaller values of $c$, and that, at a given value of $c$, stars with higher surface gravities have larger values of the scale width $b$. In the right panels of Figure~\ref{fig:BalmerLines_for_BHBiden}, we display $b$ versus $c$ for the three Balmer lines. As shown, there are clear gaps between BS stars (green dots) and known BHB stars (red dots) in all the three panels, and the gaps allow us to further refine the identification of BHB stars. For the distribution of the BHB candidates, there are clear peaks around $(c_{\beta},\ b_{\beta})=(0.94,\ 9.86\,\rm\AA)$, $(c_{\gamma},\ b_{\gamma})=(1.04,\ 9.96\,\rm\AA)$, and $(c_{\delta},\ b_{\delta})=(1.11,\ 10.04\,\rm\AA)$, and these peaks are coincident with that of the known BHB stars. In the panels using the H$\beta$ and H$\delta$ lines, there are also minor peaks at $(c_{\beta},\ b_{\beta})=(0.76,\ 9.46)$ and $(c_{\delta},\ b_{\delta})=(0.92,\ 10.60)$. Checking the LAMOST catalog indicates they are mainly late A-types. In the panel for the H$\gamma$ line, these is a long tail toward smaller values of $c_{\gamma}$, thus lower temperatures.

According to the distribution of the BHB candidates identified via the $D_{0.2}$ method, and with reference to that of the BS stars and the known BHB stars, in the $b$ versus $c$ planes, we further flag spectra meet any of the following criteria (blue lines in the right panels of Figure~\ref{fig:BalmerLines_for_BHBiden}) as BHB candidates:
{\footnotesize\begin{equation}
\begin{cases}
b_{\beta}\leq -8.66c_{\beta}^{4}+66.16c_{\beta}^{3}-167.73c_{\beta}^{2}+176.33c_{\beta}-54.81,\\
b_{\beta}\geq6.0\,{\rm \AA},\ c_{\beta}\leq1.35;\\[1.5ex]
b_{\gamma}\leq 10.92c_{\gamma}^{4}-19.14c_{\gamma}^{3}-33.96c_{\gamma}^{2}+88.25c_{\gamma}-35.05,\\
b_{\gamma}\geq6.0\,{\rm \AA},\ c_{\gamma}\leq1.45;\\[1.5ex]
b_{\delta}\leq -8.32c_{\delta}^{4}+57.45c_{\delta}^{3}-147.41c_{\delta}^{2}+165.07c_{\delta}-56.46,\\
b_{\delta}\geq6.0\,{\rm \AA},\ c_{\delta}\leq1.45.
\end{cases}
\end{equation}}

\noindent With these cuts, we flag 10\,162 spectra using the H$\beta$ line, 10\,916 spectra using the H$\gamma$ line, and 11\,269 spectra using the H$\delta$ line, as BHB candidates.

\subsubsection{The Combined Cut}\label{sec:combined_cut}

In Section~\ref{sec:Atype}, we identify a sample of A-type spectra via the $\rm EW_{G4300}$ versus $\rm EW_{H\gamma}$ plane, and in Sections~\ref{sec:D20_vs_fm} and \ref{sec:b_vs_c}, we classify these A-type spectra into BHB and non-BHB stars using the $D_{0.2}$ method and the scale width-shape method, applied to each of the three Balmer lines. There are discrepant types using different methods and among various Balmer lines. We obtain 9995, 10\,297 and 11\,108 spectra for the three Balmer lines, respectively, that both the $D_{0.2}$ method and the scale width-shape method flag as BHB candidates. In this study, we consider a spectrum to be a BHB spectrum if at least one of the three Balmer lines flags it as such\footnote{We also tried a more stringent selection strategy, i.e., only spectra flagged as BHB stars based on all three Balmer lines are included in the sample. We note that the stringent strategy will remove too many real BHB stars: for example, only half of the BHB stars from \citet{Xue2011ApJ...738...79X} satisfy this criterion. For this reason, we adopt the less stringent criterion.}. Combining the three lines, we obtain 13\,988 BHB spectra, corresponding to 10\,236 unique BHB stars. Of the BS spectra initially identified in Section~\ref{sec:D20_vs_fm}, 73 are removed due to inconsistent classifications between the two methods, and finally, we obtain 6075 BS spectra corresponding to 4282 unique BS stars. Though, we consider all spectra with $\rm SNRg\ge10$ as the parent sample (Section~\ref{sec:data_lamost}), most ($>70\%$) of the spectra in the final BHB and BS samples have $\rm SNRg\ge30$, while this fraction is $\sim55\%$ for the parent sample. The BHB stars, as well as the BS stars are provided in Table~\ref{tab:BHB_and_BS}. We include three columns (\texttt{Hbeta\_FLAG}, \texttt{Hgamma\_FLAG}, and \texttt{Hdelta\_FLAG}) in Table~\ref{tab:BHB_and_BS} to indicate individual classifications based on each of the three Balmer lines. The readers can make their own choice of the BHB sample using these columns.

\setlength{\tabcolsep}{0.75ex}\renewcommand{\arraystretch}{1.2}
\begin{longtable*}{lcp{0.70\textwidth}}
\caption[]{Properties of BHB and BS stars identified in this work}\label{tab:BHB_and_BS}\\\toprule
Column Name & Unit & Description\\[1ex]
\midrule
\endfirsthead
\caption[]{--\,{\it continued}}\\\toprule
Column Name & Unit & Description\\[1ex]
\midrule
\endhead
\bottomrule
\endfoot
\multicolumn{3}{l}{(This table is available in its entirety in machine-readable format.)}
\endlastfoot
obsid           & $\cdots$           & Unique spectrum ID number\\
RAJ2000         & deg                & Right ascension (J2000)\\
DEJ2000         & deg                & Declination (J2000)\\
SNRg            & $\cdots$           & The signal-to-noise ratio of $g$-band\\
EBV             & mag                & Color excess obtained from the extinction map of \citet{Schlegel1998ApJ...500..525S}\\
GaiaDR3         &                    & Gaia DR3 source id\\
parallax        & mas                & Parallax from Gaia DR3 (only positive parallax is shown)\\
parallax\_error & mas                & Error on parallax from Gaia DR3\\
d               & kpc                & Distance calculated as the inverse of the Gaia parallax\\
BPmag           & mag                & BP-band mean magnitude from Gaia DR3\\
e\_BPmag        & mag                & Error on BP-band mean magnitude\\
Gmag            & mag                & G-band mean magnitude from Gaia DR3\\
e\_Gmag         & mag                & Error on G-band mean magnitude\\
RPmag           & mag                & RP-band mean magnitude from Gaia DR3\\
e\_RPmag        & mag                & Error on RP-band mean magnitude\\
Jmag            & mag                & J-band magnitude from 2MASS\\
e\_Jmag         & mag                & Error on J-band magnitude from 2MASS\\
Hmag            & mag                & H-band magnitude from 2MASS\\
e\_Hmag         & mag                & Error on H-band magnitude from 2MASS\\
Hbeta\_D20      & $\rm\AA$           & The width of the H$\beta$ line at the level 20\% below the continuum level\\
Hbeta\_fm       & $\cdots$           & The minimum flux of the H$\beta$ line in the normalized spectrum\\
Hbeta\_b        & $\rm\AA$           & The scale width of the H$\beta$ line\\
Hbeta\_c        & $\cdots$           & The scale shape of the H$\beta$ line\\
Hgamma\_D20     & $\rm\AA$           & The width of the H$\gamma$ line at the level 20\% below the continuum level\\
Hgamma\_fm      & $\cdots$           & The minimum flux of the H$\gamma$ line in the normalized spectrum\\
Hgamma\_b       & $\rm\AA$           & The scale width of the H$\gamma$ line\\
Hgamma\_c       & $\cdots$           & The scale shape of the H$\gamma$ line\\
Hdelta\_D20     & $\rm\AA$           & The width of the H$\delta$ line at the level 20\% below the continuum level\\
Hdelta\_fm      & $\cdots$           & The minimum flux of the H$\delta$ line in the normalized spectrum\\
Hdelta\_b       & $\rm\AA$           & The scale width of the H$\delta$ line\\
Hdelta\_c       & $\cdots$           & The scale shape of the H$\delta$ line\\
Teff            & K                  & Effective temperature predicted by \texttt{SLAM}\\
logg            & $\rm [cm\,s^{-2}]$ & Surface gravity predicted by \texttt{SLAM}\\
FeH             & $\cdots$           & [Fe/H] predicted by \texttt{SLAM}\\
Hbeta\_FLAG     & $\cdots$           & =\texttt{BHB} for sources identified as BHB stars by the H$\beta$ line, =\texttt{BS} for sources identified as BS stars by the H$\beta$ line\\
Hgamma\_FLAG    & $\cdots$           & =\texttt{BHB} for sources identified as BHB stars by the H$\gamma$ line, =\texttt{BS} for sources identified as BS stars by the H$\gamma$ line\\
Hdelta\_FLAG    & $\cdots$           & =\texttt{BHB} for sources identified as BHB stars by the H$\delta$ line, =\texttt{BS} for sources identified as BS stars by the H$\delta$ line\\
Type            & $\cdots$           & =\texttt{BHB} for sources identified as BHB stars, =\texttt{BS} for sources identified as BS stars\\
\bottomrule
\end{longtable*}

\section{Atmospheric Parameters of BHB Stars}\label{sec:BHBatmos}

Large samples of BHB stars have been provided by various studies, either photometrically or spectroscopically \citep[e.g.,][]{Xue2008ApJ...684.1143X,Xue2011ApJ...738...79X,Ju2024ApJS..270...11J,Hu2025ApJS..278...62H}, but few studies provided systematic determination of atmospheric parameters for their BHB sample. Recently, \citet{Ju2025ApJS..276...12J} determined atmospheric parameters for a large catalog of BHB stars \citep{Ju2024ApJS..270...11J}, utilizing the \texttt{SLAM} method \citep{Zhang2020ApJS..246....9Z}. In this section, we apply the \texttt{SLAM} method to determine atmospheric parameters for our BHB sample. As a by-product, we also provide atmospheric parameters for the BS sample.

To apply the \texttt{SLAM}, we use the \texttt{normalize\_spectrum\_spline}\footnote{\url{https://laspec.readthedocs.io/en/latest/api.html\#laspec.normalization.normalize\_spectrum\_spline}} function in the PYTHON package \texttt{laspec} \citep{Zhang2020ApJS..246....9Z,Zhang2021ApJS..256...14Z} to perform spline fit to the observed spectra. This time we focus on the wavelength range of 3900\textendash5800\,$\rm\AA$, and the spectra are resampled at $1\,\rm\AA$ intervals. This wavelength range is chosen because the LAMOST low-resolution spectrum have a splicing of blue and red arms around 5700\textendash5900\;$\rm\AA$ \citep{Luo2015RAA....15.1095L}, and most of the spectral features of A-type stars are at the blue arm.

\subsection{The \texttt{SLAM} Method}\label{sec:SLAM}

The \texttt{SLAM} (Stellar LAbel Machine) is developed by \citet{Zhang2020ApJS..246....9Z}, and has been widely used in determining the stellar labels of LAMOST spectra \citep[e.g.,][]{Zhang2020RAA....20...51Z,Li2021ApJS..253...45L,Guo2021ApJS..257...54G,Ju2025ApJS..276...12J}. It is a data-driven method based on the support vector regression \citep[SVR;][]{Smola2004SVR,Chang2011SVR}, which is a robust method in modeling noisy data, and has been frequently used to analyze stellar spectra \citep[e.g.,][]{Li2014ApJ...790..105L,Liu2014RAA....14..423L,Lu2015MNRAS.452.1394L}. We briefly summarize the main points of the \texttt{SLAM} here, and the reader is referred to \citet{Zhang2020ApJS..246....9Z} for a comprehensive discussion of the algorithm.

\texttt{SLAM} has three hyperparameters, including the penalty level ($C$), the width of the radial basis function ($\gamma$), and the tube radius ($\epsilon$). Generally, there are three steps involved in applying the \texttt{SLAM} to observed spectra:

\textit{Step 1}: Preprocessing. This step is to standardize the normalized spectra in the training set, such that their spectral flux of each pixel and their stellar labels have means of 0 and variances of 1. This step is necessary for most machine learning methods, including SVR, to avoid issues due to the different scales in different dimensions of the input data.

\textit{Step 2}: Training. We feed the preprocessed training set into \texttt{SLAM} and train the SVR model with stellar labels as independent variables and spectral fluxes at each wavelength pixel as dependent variables. In this study, we train two machines: one using only flux and the other one also including color indices as additional pixels.

\textit{Step 3}: Prediction. Finally, the atmospheric parameters are predicted for our spectra applying the trained SVR model.

\begin{figure*}[!t]
    \centering
    \includegraphics[width=0.9\textwidth]{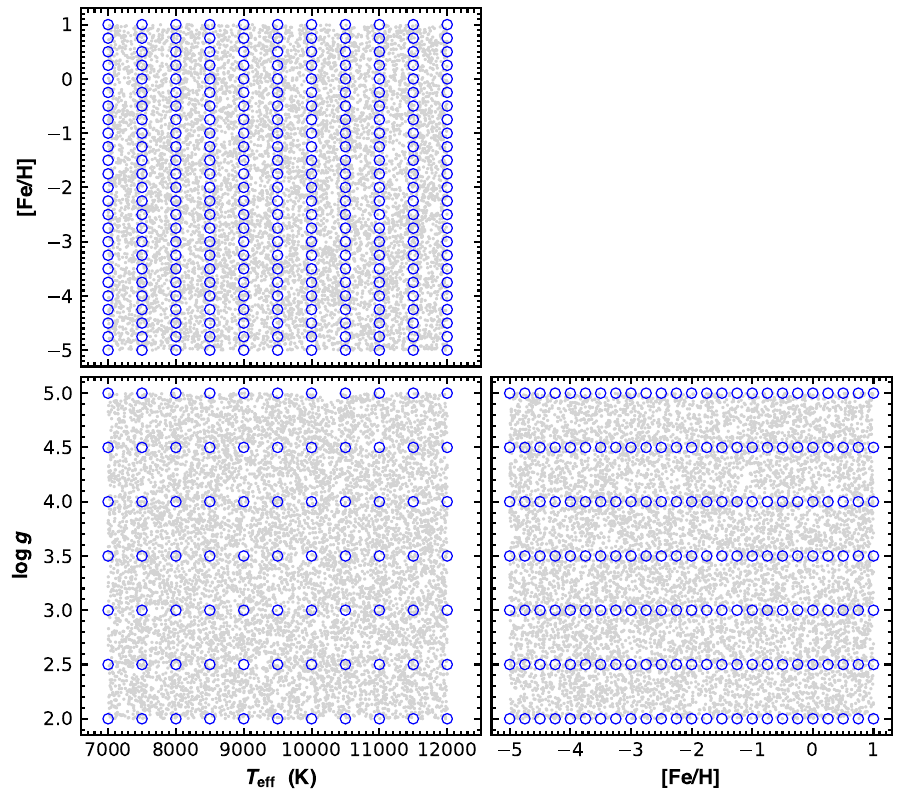}
    \caption{Corner plot showing the parameter space coverage of the theoretical spectra. The blue circles represent the spectra in the raw grid of \citet{Allende-Prieto2018A&A...618A..25A}, and the small gray dots are the linearly interpolated spectra employed for training the \texttt{SLAM}.}
    \label{fig:param_space}
\end{figure*}

\begin{figure*}[!t]
    \centering
    \includegraphics[width=\textwidth]{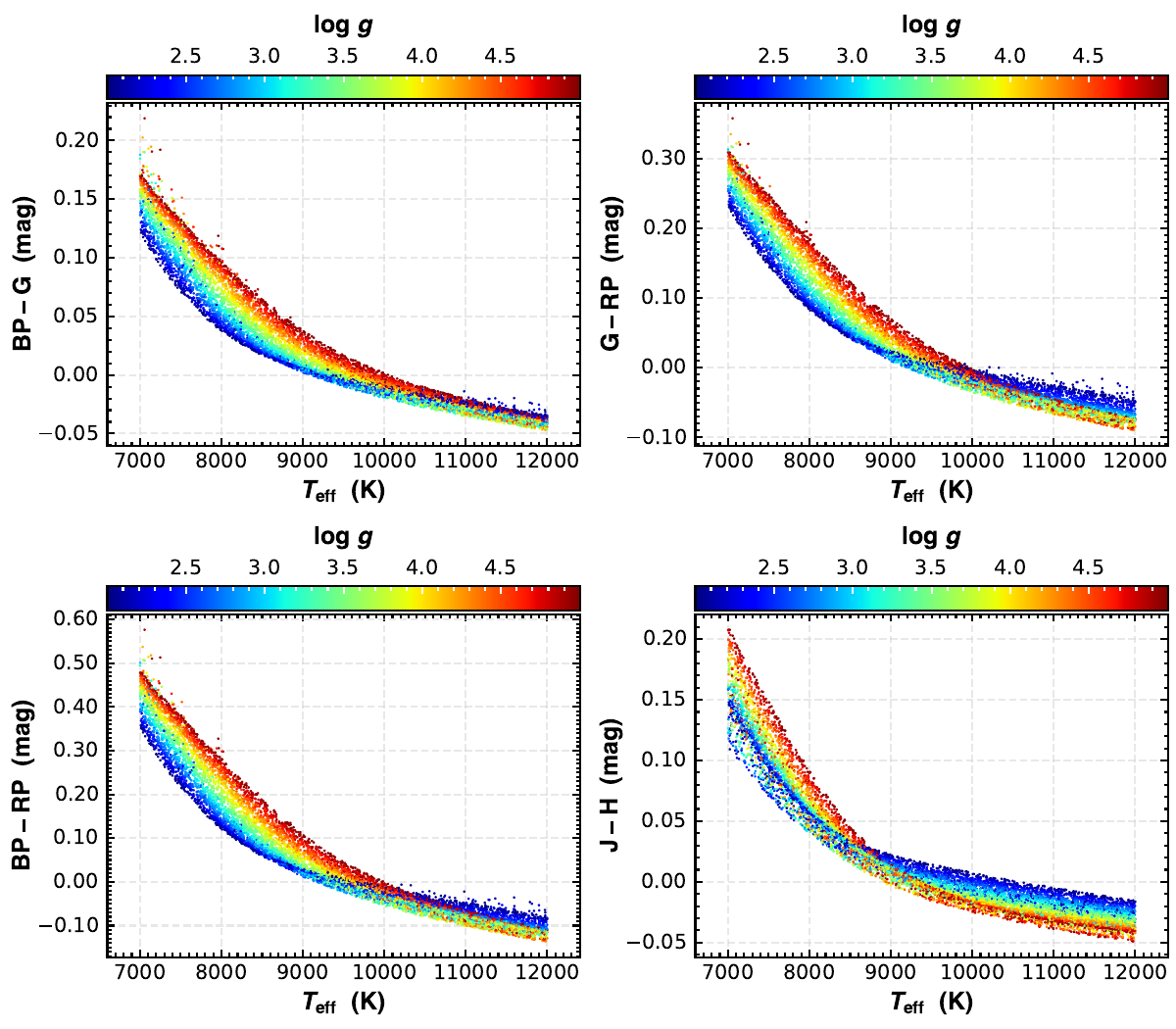}
    \caption{Intrinsic $BP-G$, $G-RP$, $BP-RP$ and $J-H$ colors as functions of effective temperatures for theoretical spectra in the training set. The points are color coded by surface gravity.}
    \label{fig:color_vs_Teff}
\end{figure*}

\subsection{The Training and Test Sets}\label{sec:trainingSet}

In order to apply the \texttt{SLAM} to the observed spectra, a training set with sufficient coverage of the parameter space is required. However, the available BHB samples cover only limited parameter space \citep{Ju2025ApJS..276...12J}, and are inadequate for our purpose. Therefore, we adopt the synthetic spectra from \citet{Allende-Prieto2018A&A...618A..25A} as the training set, which were calculated adopting the \texttt{ATLAS9} model atmospheres \citep{Meszaros2012AJ....144..120M}. \citet{Allende-Prieto2018A&A...618A..25A} provided several sets of spectral grids, with different resolutions and different sampling in the parameter space. In this study, we adopt the set with the identifier ``ns'', which contains 86\,625 spectra, with $T_{\rm eff}$ in the range 7000 to 12\,000\,K with a step of 500\,K, $\log g$ in the range 2 to 5 with a step of 0.5\,dex, [Fe/H] in the range $-5$ to 1 with a step of 0.25\,dex, [$\alpha$/Fe] in the range $-1$ to 1 with a step of 0.25\,dex, and the micro-turbulence velocity ($\log\xi$) in the range $-0.301$ to 0.903 with a step of 0.301\,dex. The spectral resolution is 10\,000 and the wavelength coverage is 2000 to 25\,000\,$\rm\AA$. In Figure~\ref{fig:param_space}, we display the model grids (blue circles) in the $(T_{\rm eff},\ \log g)$, $(T_{\rm eff},\ {\rm [Fe/H]})$, and $({\rm [Fe/H]},\ \log g)$ planes.

Since [$\alpha$/Fe] and $\log\xi$ are not relevant to this study, and cannot be well-constrained using LAMOST low-resolution spectra, we set $\log\xi=0.176$ and adopt [Fe/H]-dependent values for [$\alpha$/Fe], following the prescription used by \citet{Allende-Prieto2018A&A...618A..25A} to construct the coarse grid. This setting reduces the grid size to 1925. In order to increase the grid size, we randomly generate 10\,000 model spectra using uniformly distributed stellar parameters within the available ranges through linear interpolations, and the new grids are shown as gray dots in Figure~\ref{fig:param_space}.

We also generate another set of 2000 spectra to test the performance of the \texttt{SLAM}, and study the trend with noise levels.

The theoretical spectra are convolved to the LAMOST resolution (i.e., $R\approx1800$), and resampled at similar wavelength points as the LAMOST observation. Then, the convolved and resampled theoretical spectra are normalized as done for the observed spectra.

We should mention that while the Balmer lines are sensitive indicators of effective temperatures for A-type stars, they are not monotonic functions of effective temperature, instead they peak around spectral type A2 ($T_{\rm eff}\approx8500\textendash9000\,\rm K$; see \citealt{Fairlamb2017MNRAS.464.4721F} for an example). There are degeneracy between effective temperature and surface gravity, making hot giants and cool dwarfs displaying similar Balmer line profiles. Color indices provide another measurements of effective temperature that are less impacted by the degeneracy with surface gravities \citep[e.g.,][]{Allende-Prieto2006ApJ...636..804A, Ju2025ApJS..276...12J}. For this reason, synthetic photometry and colors are also obtained for the theoretical spectra with the PYTHON package \texttt{PYPHOT} \citep{Fouesneau2025zndo..14712174F_PYPHOT}. In Figure~\ref{fig:color_vs_Teff}, we display the synthetic $BP-G$, $G-RP$, $BP-RP$, and $J-H$ colors as functions of effective temperature. As shown, all the four color indices decrease monotonically with increasing temperature, while the impact of surface gravity is minor. These color indices are appended to the flux pixels to train the \texttt{SLAM} model.

\subsection{The Performance of the SLAM}\label{sec:performance}

\begin{figure*}[!t]
    \centering
    \includegraphics[width=\textwidth]{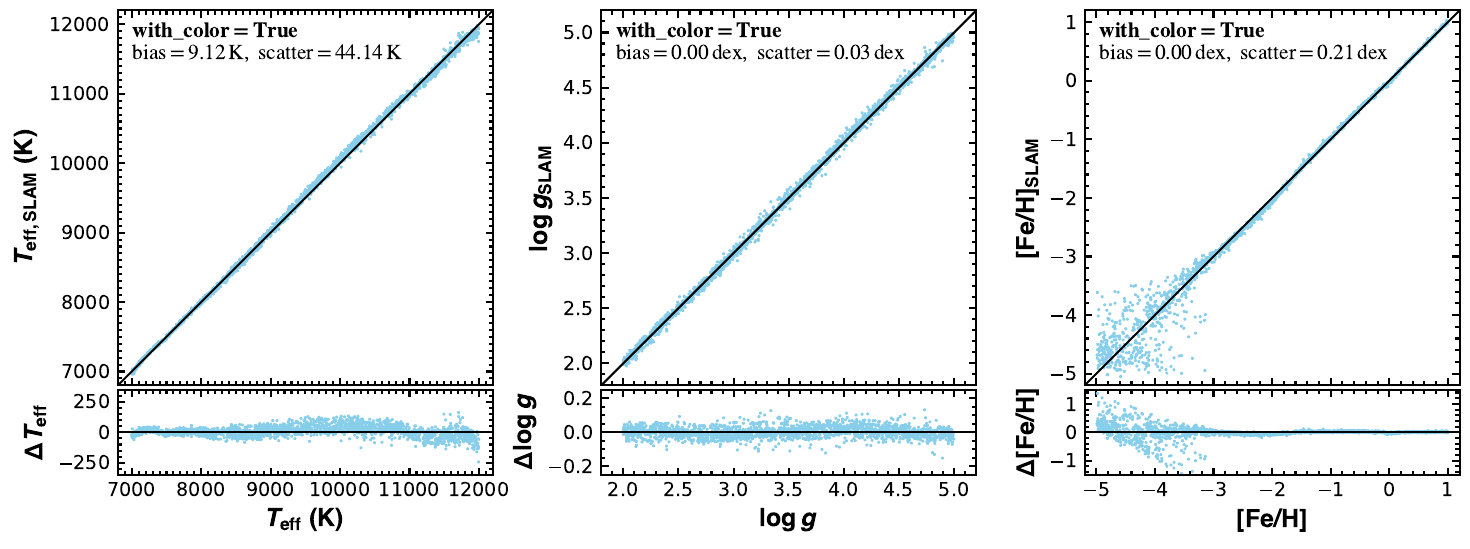}
    \caption{Comparison between the predicted labels and the true labels for the test set when \texttt{SLAM} is trained with both flux and colors. The black lines in the upper panels are the line of equality. The bottom panels display the residuals. The CV-scatter and CV-bias values are labeled.}
    \label{fig:comp_SLAM1}
\end{figure*}

\begin{figure*}[!t]
    \centering
    \includegraphics[width=\textwidth]{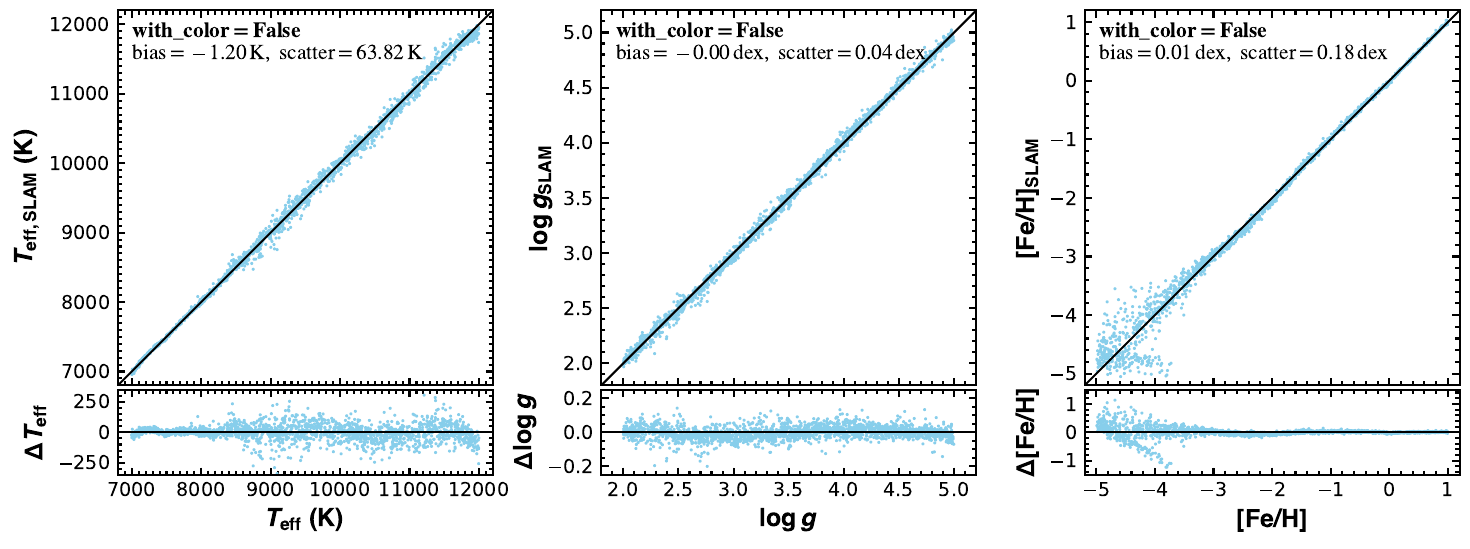}
    \caption{Similar as Figure~\ref{fig:comp_SLAM1}, but the \texttt{SLAM} is trained with only flux.}
    \label{fig:comp_SLAM0}
\end{figure*}

Before applying \texttt{SLAM} to predict stellar labels of the observed spectra, it is necessary to verify the validity of the machine. To achieve this, we first apply the trained machines to the test set. Following \citet{Zhang2020ApJS..246....9Z}, we define the cross valid scatter (CV-scatter) and the cross valid bias (CV-bias) as follows:
\begin{equation}
\text{CV-scatter}=\sqrt{\dfrac{1}{n}\sum_{i=1}^{n}\left(\theta_{i,\rm SLAM}-\theta_{i}\right)^{2}},
\end{equation}
and
\begin{equation}
\text{CV-bias}=\dfrac{1}{n}\sum_{i=1}^{n}\left(\theta_{i,\rm SLAM}-\theta_{i}\right),
\end{equation}
where $n$ is the sample size of the test set (i.e., $n=2000$ here), $\theta_{i}$ is the true label of the $i$th spectrum in the test set, and $\theta_{i,\rm SLAM}$ denote the \texttt{SLAM} predicted label for the same spectrum. In Figures~\ref{fig:comp_SLAM1} and \ref{fig:comp_SLAM0}, we display the comparison between the predicted labels and the true labels, when \texttt{SLAM} is trained with and without color information, respectively. The scatter values are 42\,K, 0.03\,dex, and 0.22\,dex for $T_{\rm eff}$, $\log g$, and [Fe/H], respectively, when training \texttt{SLAM} with both flux and colors, and the values are 58\,K, 0.04\,dex, and 0.19\,dex when training \texttt{SLAM} without colors. The bias values are negligible for both machines, indicating that there are no systematic bias. We note that the CV-scatter values are significantly larger for [Fe/H] than for $\log g$, this is caused mainly by the spectra with low metallicities, since there are few metal lines in these metal-poor spectra that can be used to derive [Fe/H] accurately. Indeed, when only spectra with $\rm [Fe/H]>-3$ are considered, the scatter values decrease to 0.04\textendash0.05\,dex.

\begin{figure*}[!t]
    \centering
    \includegraphics[width=\textwidth]{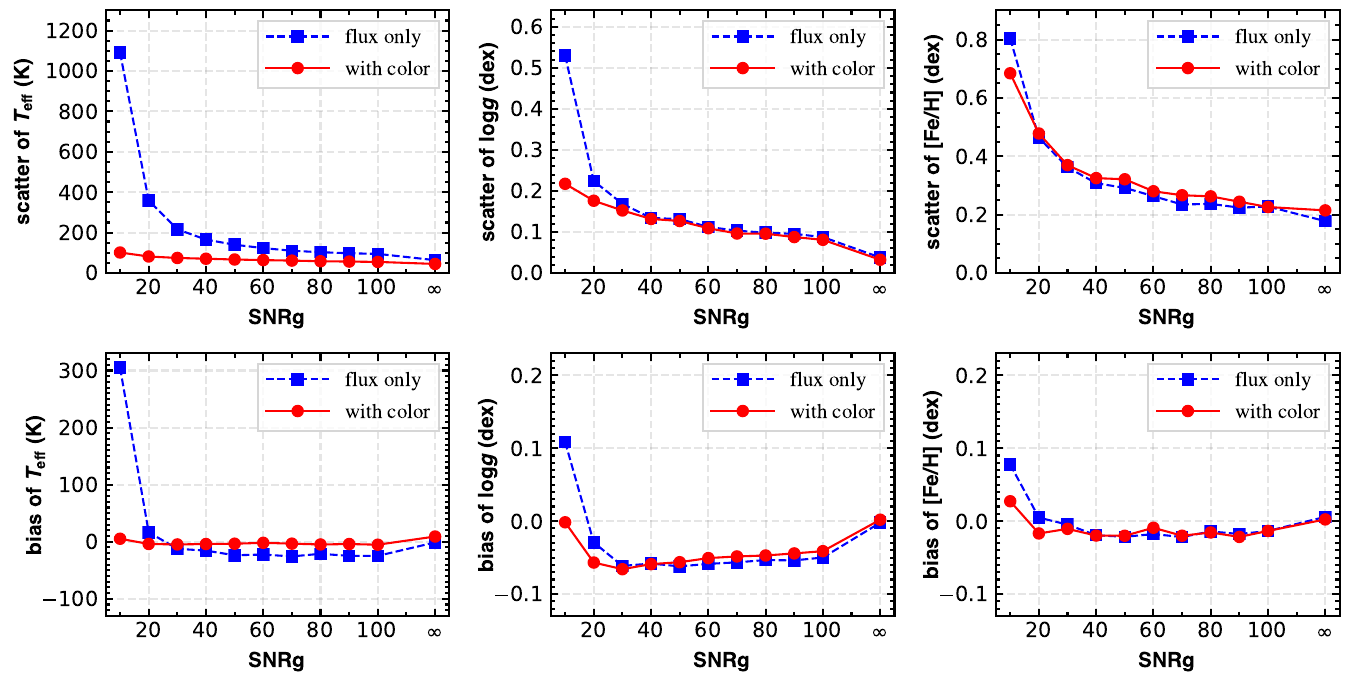}
    \caption{The distribution of the CV-scatter (top panels) and CV-bias (bottom panels) values for predicted stellar labels of $T_{\rm eff}$ (left panels), $\log\,g$ (middle panels), and [Fe/H] (right panels) as functions of SNRg values. In each panel, blue squares connected with dashed line represent training \texttt{SLAM} with flux only, and red circles connected with solid line represent training \texttt{SLAM} with both flux and colors. The rightmost points in each panel represent the scatter and bias values for labeling the noise-free ($\text{SNRg}=\infty$) spectra in the test set.}
    \label{fig:CVvalue_vs_snr}
\end{figure*}

\begin{figure*}[!t]
    \centering
    \includegraphics[width=\textwidth]{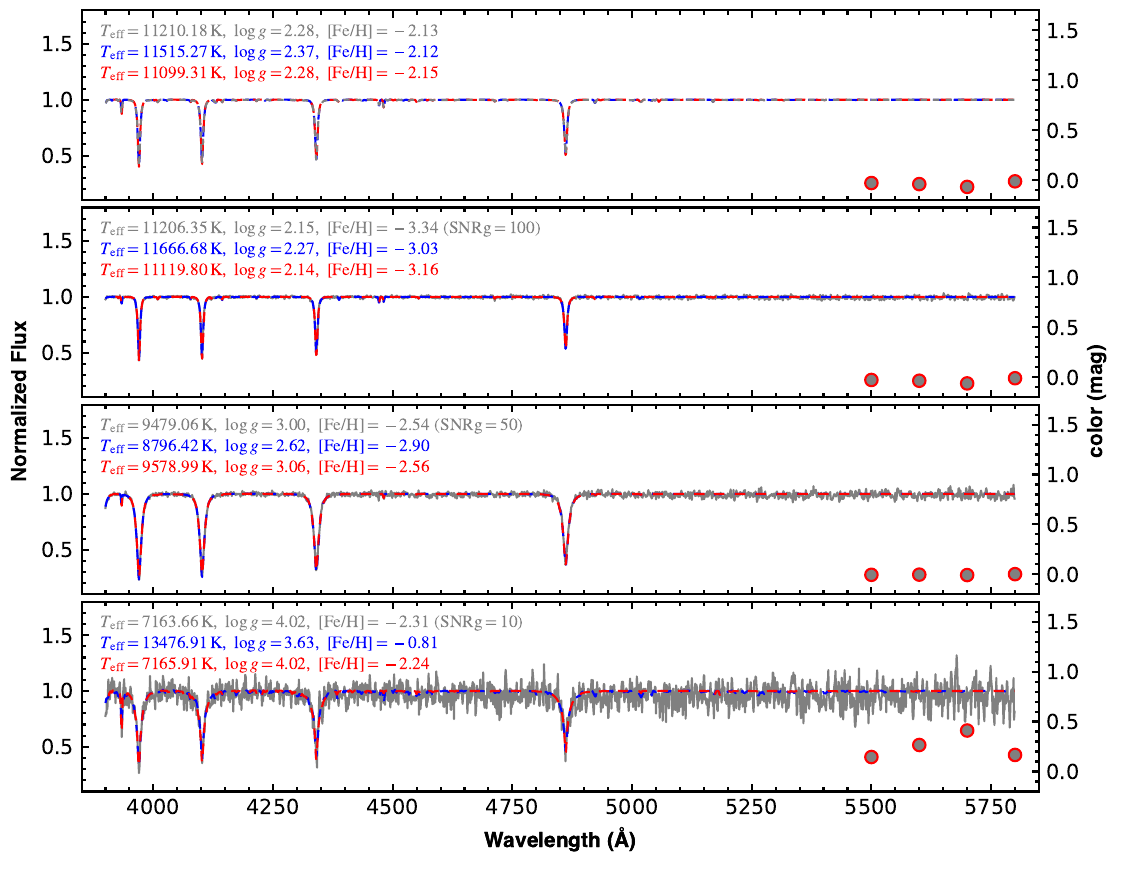}
    \caption{Top-most panel: An example showing the fits to a noise-free spectrum in the test set. The test spectrum is shown as gray line, and the corresponding $BP-G$, $G-RP$, $BP-RP$, and $J-H$ colors are indicated with gray dots from left to right (see the right y-axis). The blue line is the \texttt{SLAM} predicted spectrum when training \texttt{SLAM} with only flux. The red line and circles represent the \texttt{SLAM} predicted spectrum and colors, respectively, when training \texttt{SLAM} with both flux and color indices. The true labels and the predicted labels are indicated with the corresponding colors. The other three panels display fits to test spectra adding different levels of gaussian noise.}
    \label{fig:SPECexample_SLAMtheo}
\end{figure*}

\begin{figure*}[!t]
    \centering
    \includegraphics[width=\textwidth]{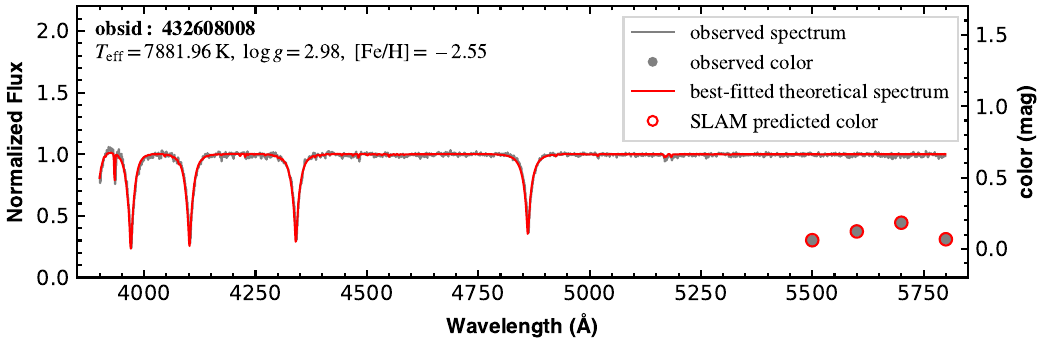}
    \caption{An example demonstrating fitting the observed spectrum with the \texttt{SLAM} trained with both flux and color indices. The observed and the best-fitted spectrum are shown as gray and red lines, respectively. The gray dots and red circles represent the observed and the \texttt{SLAM} predicted colors, respectively. From left to right are $BP-G$, $G-RP$, $BP-RP$, and $J-H$ colors, respectively (see the right y-axis).}
    \label{fig:SPECexample_SLAMobs}
\end{figure*}

\begin{figure*}[!t]
    \centering
    \includegraphics[width=\textwidth]{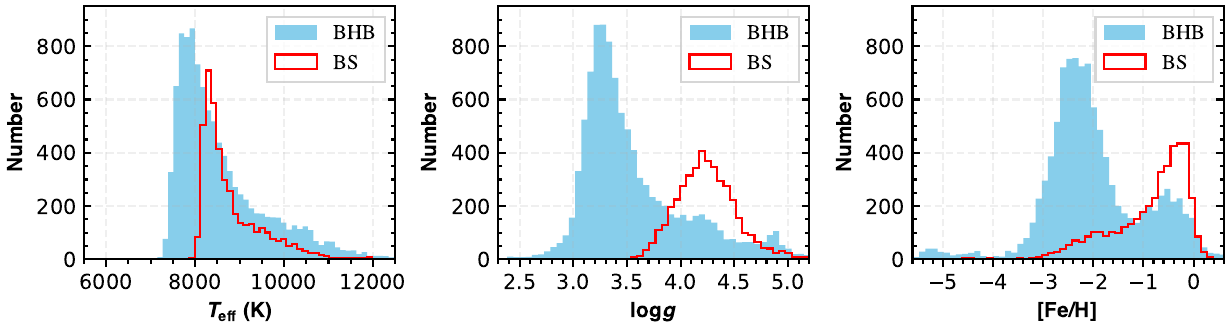}
    \caption{Histograms showing the distribution of $T_{\rm eff}$ (left panel), $\log g$ (middle panel), and [Fe/H] (right panel) for the BHB (blue solid) and BS (red open) samples.}
    \label{fig:HIST_params}
\end{figure*}

\begin{figure*}[!t]
    \centering
    \includegraphics[width=0.8\textwidth]{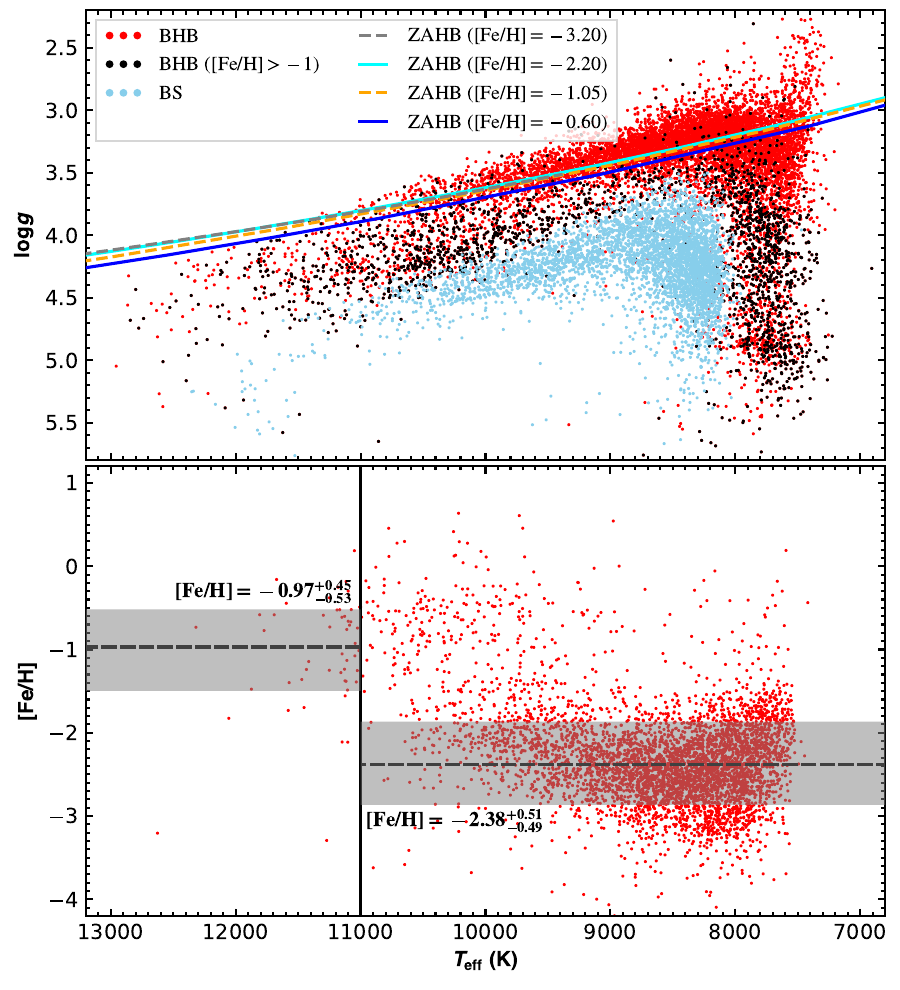}
    \caption{Top panel: Distribution of BHB (red and black dots) and BS (blue dots) stars in the $\log g$ vs. $T_{\rm eff}$ plane. Also shown are the theoretical zero-age horizontal branches (ZAHBs) with various [Fe/H] from \citet{Pietrinferni2021ApJ...908..102P}. BHB stars with $\rm [Fe/H]>-1$ are shown as black dots. Bottom panel: Distribution of BHB stars in the [Fe/H] vs. $T_{\rm eff}$ plane, and only objects classified as BHB stars based on all the three Balmer lines are included in the plot. The vertical solid line corresponds to $T_{\rm eff}=11\,000\,{\rm K}$, dividing the sample into hot and cool subsamples. The horizontal dashed lines show the median [Fe/H] values of the hot and cool subsamples, respectively, and the gray shaded bands display the corresponding 1$\sigma$ ranges. The median and 1$\sigma$ values are labeled as well.}
    \label{fig:Kiel_diagram}
\end{figure*}

The CV-scatter and CV-bias values are also dependent on the signal-to-noise ratios of the input spectra. We thus explore the trend of CV values as functions of SNRg. We add gaussian noise with a set of SNRg to each pixel of the spectrum in the test set. Then we apply the trained \texttt{SLAM} model to the noisy spectra to evaluate the performance of \texttt{SLAM}. In this work, we choose SNRg values range from 10 to 100 with an increment of 10.

In Figure~\ref{fig:CVvalue_vs_snr}, we show the distribution of CV-scatter and CV-bias values as functions of SNRg. As shown, there are clear trend of decreasing CV values with increasing SNRg values, indicating that the data-driven method is suitable. The bias is always negligible when $\rm SNRg\ge20$, further indicating that the model is applicable to our spectra, since the majority of our spectra have $\text{SNRg}\ge20$. The predicted labels are less biased and exhibit smaller scatters when training \texttt{SLAM} with both flux and colors than when training \texttt{SLAM} with only flux, and the improvement is especially pronounced for low SNRg spectra. We also note that the improvement is more significant for the estimation of $T_{\rm eff}$ and $\log g$ than for that of [Fe/H]. This is to be expected since a color index is an effective temperature metric and can help break the degeneracy between $T_{\rm eff}$ and $\log g$, but carries little information on metallicity.

Figure~\ref{fig:SPECexample_SLAMtheo} shows several examples further demonstrating the necessity of including color indices in the training process. When SNRg is high, the results of training \texttt{SLAM} with and without color indices are comparable to each other, and both match the true values fairly well. But for low SNRg spectra, there are significant difference between the two results. As shown in the bottom-most panel, the noisy spectrum of a hot giant may mimic that of a cool dwarf, making determination of its parameters highly uncertain when relying solely on spectra. But including color indices in the fitting can effectively break the degeneracy between $T_{\rm eff}$ and $\log g$, and the determined parameters match the true labels excellently well.

\subsection{Applying the \texttt{SLAM} to Observed Spectra}\label{sec:BHB_with_SLAM}

In the previous sections, we trained the \texttt{SLAM} and demonstrated that the \texttt{SLAM} trained with both flux and colors outperforms that trained with only flux. In this section, we apply the machine trained with both flux and colors to the observed spectra for the BHB sample. As a by-product, we also determine the atmospheric parameters for the BS sample. A spectrum example is shown in Figure~\ref{fig:SPECexample_SLAMobs}, and the full catalog for both BHB and BS stars is provided in Table~\ref{tab:BHB_and_BS}. Distribution of the predicted stellar labels are shown in Figure~\ref{fig:HIST_params}. The median values of the BHB stars are 8270\,K for $T_{\rm eff}$, 3.41\,dex for $\log g$, and $-2.16$\,dex for [Fe/H], consistent with literature values \citep{Santucci2015ApJ...801..116S,Bystrom2025MNRAS.542..560B,Ju2025ApJS..276...12J}. Compared with the distribution for BS stars, BHB stars have systematically lowered [Fe/H] and $\log g$ values. We also note a bump at $\rm [Fe/H]\approx-0.5$ in the distribution of [Fe/H], and we will discuss it further in Section~\ref{sec:dis:bump}. In the top panel of Figure~\ref{fig:Kiel_diagram}, we display the distribution of BHB and BS stars in the $\log g$ versus $T_{\rm eff}$ plane (i.e., the Kiel diagram). As shown, the BHB stars form a well defined narrow strip, and BHB stars display smaller $\log g$ values than BS stars with similar $T_{\rm eff}$. When compared to the zero-age horizontal branches (ZAHBs) taken from \citet{Pietrinferni2021ApJ...908..102P}, the distribution of the BHB stars matches the theoretical trend fairly well. We note there is a clump of stars with $T_{\rm eff}$ between 7500 to 8000\;K and $\log g$ higher than $\sim4.0$ in the plane. Many of these stars display discrepant types among various Balmer lines, and more than half ($\sim55\%$) of these stars have $\rm [Fe/H]\ge-1$, while only $\sim15\%$ of other stars do. Though these stars deviate significantly from the ZAHBs, and may suffer higher contamination than the other stars, they are included in our sample for the sake of completeness.

Discontinuities along the blue HBs are commonly observed in globular cluster, such as the ``Grundahl jump" (G-jump) at $T_{\rm eff}\sim11\,500\pm500\,{\rm K}$ \citep{Grundahl1998ApJ...500L.179G,Grundahl1999ApJ...524..242G}, and the ``Momany jump'' (M-jump) at $T_{\rm eff}\sim23\,000\,{\rm K}$ \citep{Momany2002ApJ...576L..65M}. In this work, we focus on the HBA sub-category that is cooler than 12\,000\,K, so we briefly discuss the G-jump. To minimize the contamination from BS stars, we consider only objects that are classified as BHB stars based on all the three Balmer lines. This high-purity subsample is plotted in the [Fe/H] versus $T_{\rm eff}$ plane in the bottom panel of Figure~\ref{fig:Kiel_diagram}. As shown, there exists a jump around $T_{\rm eff}\approx11\,000\,{\rm K}$, consistent with the commonly quoted G-jump value \citep[e.g.,][]{Grundahl1999ApJ...524..242G,Latour2023A&A...677A..86L}, and hot BHB stars ($T_{\rm eff}\ge11\,000\,{\rm K}$) are systematically more metal-rich than cool ones ($T_{\rm eff}<11\,000\,{\rm K}$). The median [Fe/H] values are $-0.97$ and $-2.38\,{\rm dex}$ for hot and cool BHB stars, respectively.

\begin{figure*}[!t]
    \centering
    \includegraphics[width=\textwidth]{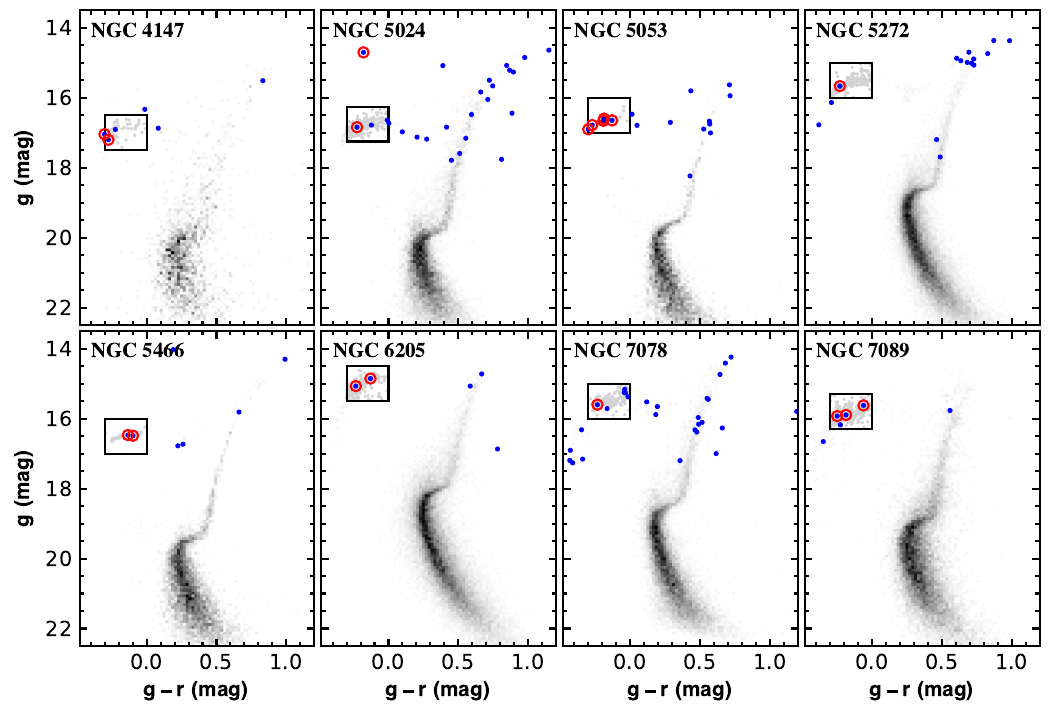}
    \caption{CMDs of 8 Globular clusters in the SDSS footprint \citep{An2008ApJS..179..326A}. In each panel, the background image shows SDSS photometry for the corresponding cluster. The black boxes show the criteria used to select BHB stars in the clusters following \citet{Vickers2012AJ....143...86V}, and the selected BHB stars are shown as gray dots. The blue dots mark sources observed with the LAMOST, and sources identified as BHB stars in this work are highlighted with additional red circles. All the stars displayed are dereddened using the extinction map of \citet{Schlegel1998ApJ...500..525S} and the extinction law of \citet{Wang2019ApJ...877..116W}.}
    \label{fig:CMD_of_GCs}
\end{figure*}

\begin{figure*}[!t]
    \centering
    \includegraphics[width=\textwidth]{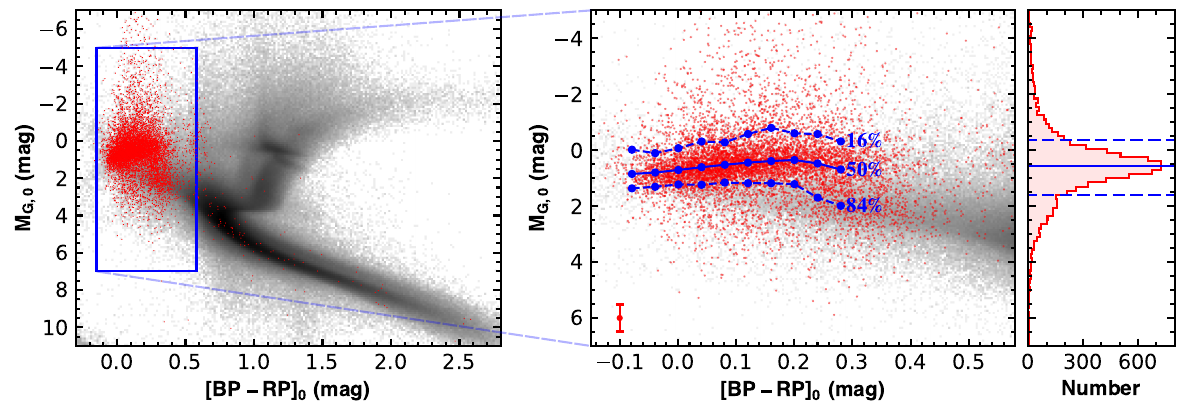}
    \caption{Left panel: Distribution of BHB stars (red dots) in the Gaia CaMD. The background image shows the distribution of all the LAMOST sources with $|b|\ge20^{\circ}$ and $\rm SNRg\ge10$ (see Section~\ref{sec:data_lamost}). The extinction correction is performed adopting the extinction map of \citet{Schlegel1998ApJ...500..525S} and the extinction law of \citet{Wang2019ApJ...877..116W}. Middle panel: Zoom-in view of the horizontal-branch (blue box in the left panel). The solid line shows the median trend of $M_{G,0}$ as a function of $[BP-RP]_{0}$ for the BHB stars. The dashed lines represent the trend of the 16th and 84th percentiles. The errorbar in the lower left corner shows the typical uncertainty of $M_{G,0}$ ($\sim$0.5\;mag, and see the text for details). Right panel: Histogram showing the distribution of $M_{G,0}$. The solid line shows the median value of $M_{G,0}$, and the two dashed lines mark the 16th and 84th percentiles.}
    \label{fig:BHB_on_CMD}
\end{figure*}

\section{Discussion}\label{sec:dis}
\subsection{Comparison with Other BHB Samples and Assessment of Sample Contamination}\label{sec:dis:Com_and_Con}

\citet{Ju2024ApJS..270...11J} identified 5436 BHB stars based on LAMOST DR5, by analyzing the Balmer line profiles as well. Of these BHB stars, 5250 have matches with $\rm SNRg\ge10$ in LAMOST DR11. The unmatched stars are mainly due to the update of the LAMOST pipeline, rendering some spectra in LAMOST DR5 are not identical to that in LAMOST DR11, or are missing. Of the 5250 matches in LAMOST DR11, 4593 ($\sim$90\%) are identified as BHB stars in this work as well. Also based on LAMOST DR5, \citet{Vickers2021ApJ...912...32V} identified 13\,693 BHB spectra using the machine learning method. Of these BHB spectra, 6989 have $|b|\geq20^{\circ}$, and 6617 of them also have matches in LAMOST DR11 and $\rm SNRg\geq10$. Of the 6617 matches, 5018 ($\sim$80\%) are identified as BHB spectra by us. Combining the photometric selection, the $D_{0.2}$ method, and the scale width-shape method, \citet{Xue2011ApJ...738...79X} identified 4985 BHB stars from SDSS DR8 \citep{Aihara2011ApJS..193...29A}. Of these stars, 1730 have matches in the parent sample (see Section~\ref{sec:data_lamost}), and 1433 ($\sim$80\%) are identified as BHB stars. Recently, \citet{Culpan2024A&A...685A.134C} presented a Gaia DR3-based catalog of 9172 BHB candidates with atmospheric and stellar parameters calculated from synthetic SEDs. Of these BHB candidates, 1494 objects have matches in the parent sample (see Section~\ref{sec:data_lamost}), and 1338 ($\sim$90\%) are identified as BHB stars by us. Based on these comparisons, we estimate an identification rate $\sim$80\%\textendash90\% for our method.

We use globular clusters (GCs) to further investigate the identification rate of our method. For this purpose, we collect 8 GCs from the SDSS footprint \citep{An2008ApJS..179..326A} and use the same selection boxes as \citet{Vickers2012AJ....143...86V} to identify the horizontal-branch of the clusters. Extinction corrected color-magnitude diagrams (CMDs) for these clusters are shown in Figure~\ref{fig:CMD_of_GCs}. Though very few sources are observed with the LAMOST, nearly all the BHB stars observed by the LAMOST are correctly identified in this work. The identification rate is low in the clusters NGC~5024 and NGC~7078, which is mainly due to that the selection box suffer RR Lyrae contamination on the red end \citep{Vickers2012AJ....143...86V}\footnote{There are 7 objects toward NGC~5024 and NGC~7078 that fall in the selection boxes, and are observed with LAMOST, but are not identified as BHB stars by us. Five of them are classified as RR Lyrae variables by various studies \citep[i.e.,][]{Arellano-Ferro2012MNRAS.420.1333A,Drake2014ApJS..213....9D,Siegel2015AJ....150..129S}. For the remaining two objects, one is classified as an RR Lyrae variable based on the variability weight and variability evidence criteria defined in \citet{Stetson2019MNRAS.485.3042S}, and the other one display variability evidence just below the threshold of 2.}. Excluding these two clusters, there are 16 BHB stars in the other 6 clusters observed by LAMOST, and 13 of them are correctly identified as BHB stars in this work, resulting in a rate of $\sim$81\%, that is consistent with the rate in the previous comparisons (i.e., 80\%\textendash90\%).

In this work, we identify 18 BHB stars in the 8 GCs, and 15 of them are within the selection boxes, indicating a contamination rate of $\sim$17\%. Of the three stars outside the selection boxes, only the star toward NGC~5024 deviates significantly from the selection box, so the contamination rate is even lower ($\lesssim10$\%). To further investigate the purity of the BHB sample, we place them in the extinction corrected color-absolute magnitude diagram (CaMD). As shown in Figure~\ref{fig:BHB_on_CMD}, they form a well defined horizontal branch, with 1$\sigma$ scatter of $\sim$0.8\,mag for $M_{G,0}$. There are four contributors responsible for the scatter in $M_{G,0}$, including the intrinsic scatter of the absolute magnitudes of BHB stars, the photometric uncertainty on the apparent magnitudes, the extinction correction, and the uncertainty on the parallax measurements. The intrinsic scatter of $M_{G,0}$ is $\sim$0.3\,mag\footnote{This value is estimated using the 8 GCs shown in Figure~\ref{fig:CMD_of_GCs}. We determine extinction corrected apparent magnitude in the Gaia $G$-band for the BHB stars in each GC, and the 1$\sigma$ scatter values are in the range of 0.1 to 0.5\,mag, with typical value of $\sim$0.3\,mag.}. The photometric uncertainty on the Gaia $G$-band is very small (mostly $<$0.01\,mag). Uncertainty due to extinction correction is also very small, since the extinction values ($A_{V}$) obtained from the extinction map of \citep{Schlegel1998ApJ...500..525S} are generally $<$0.2\,mag. Uncertainty results from the parallax measurements is $\sim$0.5\,mag, significantly larger than the other contributors, since Gaia parallaxes become less precise for distant BHB stars. Combining these contributors together results in a value of $\sim$0.6\,mag, that is comparable to the observed scatter, indicating that most of the stars are genuine BHB stars.

\subsection{Spatial Distribution of the BHB Stars}\label{sec:dis:BHBspat}

BHB stars are known to have nearly constant absolute magnitudes, and thus, their distances can be measured by comparing the observed magnitudes to their intrinsic absolute magnitudes (e.g., $M_{g}\approx0.7\,\rm mag$; \citealt{Yanny2000ApJ...540..825Y}). This method has been applied to measure distances of BHB stars in the Galactic halo with an accuracy of $\sim$10\% \citep{Sirko2004AJ....127..899S,Xue2008ApJ...684.1143X}.

\begin{figure*}[!t]
    \centering
    \includegraphics[width=0.9\textwidth]{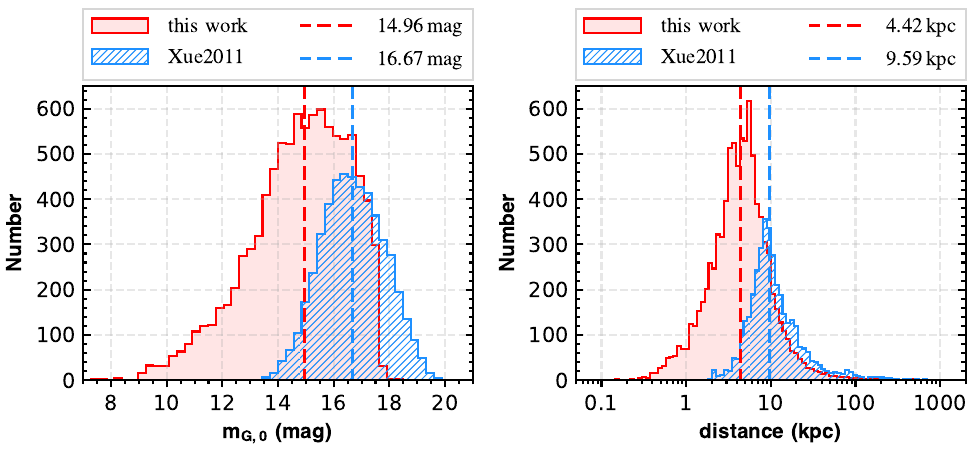}
    \caption{Histograms showing the distribution of apparent magnitude (left panel) and distance (right panel) of our BHB sample (red filled histograms) and the BHB sample from \citet{Xue2011ApJ...738...79X} (blue hatched histograms). The red and blue dashed lines show the median values of apparent magnitude and distance of the two samples, respectively, and the corresponding values are indicated in the legends.}
    \label{fig:BHB_hist_mag_dist}
\end{figure*}

The left panel of Figure~\ref{fig:BHB_hist_mag_dist} compares the distribution of the extinction corrected apparent magnitudes in the Gaia $G$-band of BHB stars in this work and \citet{Xue2011ApJ...738...79X}, where the correction is performed adopting the extinction map of \citet{Schlegel1998ApJ...500..525S} and the extinction law of \citet{Wang2019ApJ...877..116W}. The median values are 14.96\,mag and 16.67\,mag for our sample and that from \citet{Xue2011ApJ...738...79X}, respectively. Our BHB stars are brighter and thus closer than that in \citet{Xue2011ApJ...738...79X}. The right panel of Figure~\ref{fig:BHB_hist_mag_dist} compares the distance distributions of the two samples. Though the conversion from Gaia parallax to distance is highly uncertain for these distant stars, the difference between the median distance of the two samples matches the magnitude difference fairly well. Both comparisons indicate that the BHB stars in this work are much closer than that in \citet{Xue2011ApJ...738...79X}.

\begin{figure}[!t]
    \centering
    \includegraphics[width=\columnwidth]{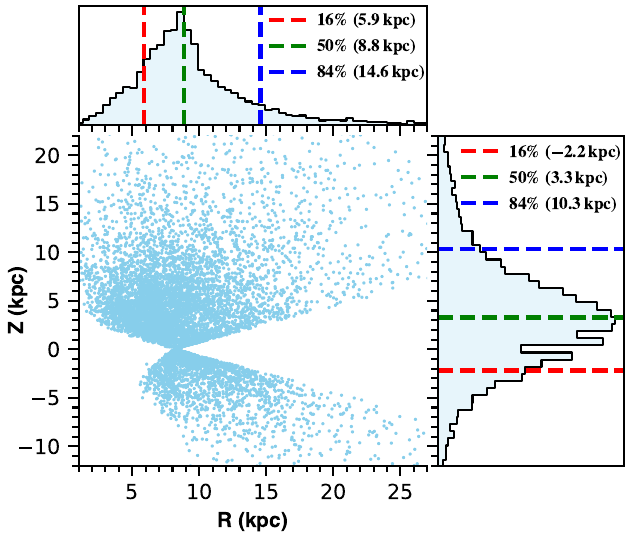}
    \caption{Corner plot showing the spatial distribution of the BHB stars in the $R\textendash Z$ plane of the cylindrical galactocentric coordinate system. The red, green, and blue dashed lines in the diagonal panels show the 16th, 50th, and 84th percentiles for $R$ and $Z$, respectively, and the corresponding values are indicated in the legends.}
    \label{fig:BHBspat}
\end{figure}

Figure~\ref{fig:BHBspat} displays the distribution of BHB stars in the $R\textendash Z$ plane of the cylindrical galactocentric coordinate system. As shown, most ($\sim$80\%) of the BHB stars are within 10\;kpc from the galactic plane, and the distribution peaks at 3\textendash4\,kpc above the galactic plane. Unlike the studies of \citet{Xue2008ApJ...684.1143X,Xue2011ApJ...738...79X}, which were focus on the halo population, our sample contains $\sim$1000 BHB stars located 1\textendash2\;kpc from the galactic plane, and these stars are likely to belong to the thick disk of the Galaxy \citep{Beers2002AJ....124..931B}. Due to their large brightness, these thick disk stars are good tracers for studying the thick disk structure, and testing theories about the origin of the thick disk \citep[e.g.,][]{Pinna2024A&A...683A.236P}. Our sample also contains BHBs close to the Galactic plane ($|Z|<1\,\rm kpc$) and in the solar vicinity. They are good tracers for studying the Galactic structure in the solar vicinity.

\subsection{Comparing the Atmospheric Parameters with the BHB Sample of \citet{Culpan2024A&A...685A.134C}}\label{sec:dis:C24}
\begin{figure*}[!t]
    \centering
    \includegraphics[width=0.9\textwidth]{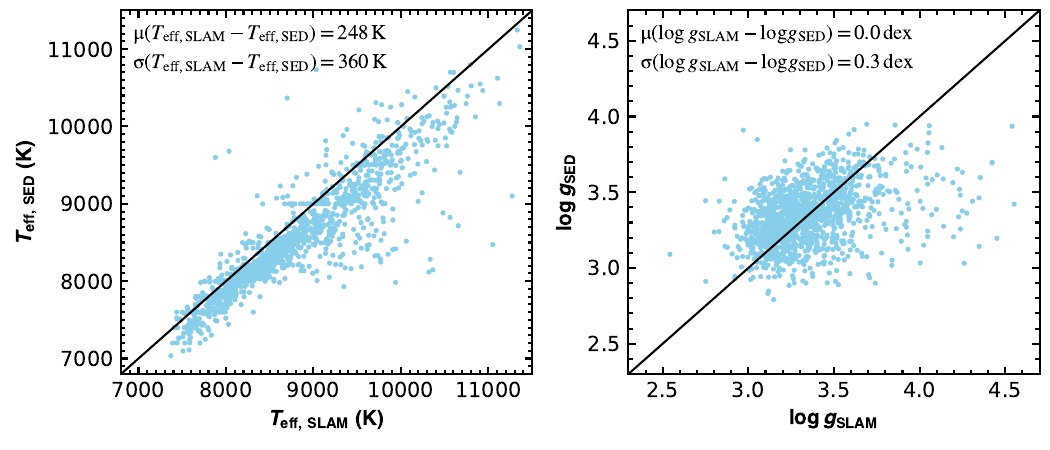}
    \caption{Comparison of $T_{\rm eff}$ (left panel) and $\log g$ (right panel) obtained in this work with that from \citet{Culpan2024A&A...685A.134C}. In each panel, the solid line is the line of equality. Mean and standard deviation of the parameter differences between our results and that from \citet{Culpan2024A&A...685A.134C} are also shown in the plot.}
    \label{fig:parsHere_vs_parsC24}
\end{figure*}

\citet{Culpan2024A&A...685A.134C} estimated atmospheric parameters for a sample of over 9000 BHB stars through spectral energy distribution (SED) analysis. The SED-based method is completely different from the \texttt{SLAM} method adopted in this work. We compare the atmospheric parameters of \texttt{SLAM} with theirs in Figure~\ref{fig:parsHere_vs_parsC24}. Only $T_{\rm eff}$ and $\log g$ are compared, since the metallicity can not be well determined from the SED alone. As shown in the left panel, the \texttt{SLAM}-predicted $T_{\rm eff}$ values are systematically higher by $\sim250\,{\rm K}$ than the SED-based values. We note that the theoretical spectra used in this work were computed with the \texttt{ATLAS9} code \citep{Kurucz1979ApJS...40....1K}, while the synthetic SEDs used in \citet{Culpan2024A&A...685A.134C} were calculated with the \texttt{ATLAS12} code \citep{Irrgang2018A&A...615L...5I}. We attribute the systematic shift between the \texttt{SLAM}-predicted and SED-based $T_{\rm eff}$ to the intrinsic differences between the theoretical grids adopted in this work and those used by \citet{Culpan2024A&A...685A.134C}\footnote{We performed a simple test using the \texttt{ATLAS9}-based spectra to fit the \texttt{ATLAS12}-based spectra, and the fitted $T_{\rm eff}$ values are $\sim150\,{\rm K}$ higher than the reference values. This trend is similar to the one observed.}. The $\log g$ values are compared in the right panel of Figure~\ref{fig:parsHere_vs_parsC24}, and there are no systematic differences between the \texttt{SLAM}-predicted values and the SED-based values.

\subsection{The Bump Around [Fe/H]=-0.5}\label{sec:dis:bump}

While BHB stars are generally metal-poor stars with typical values of [Fe/H] in the range $-3$ to $-1$\,dex \citep{Kinman2000A&A...364..102K}, relatively metal-rich BHB stars have been found by various studies. Using high-resolution optical spectra, \citet{Behr1999ApJ...517L.135B} found the iron of hot BHB stars ($T_{\rm eff}>12\,000\,K$) in the globular cluster M13 is enhanced to 3 times the solar abundance, 2 orders of magnitude above the canonical metallicity of $\rm [Fe/H]\approx-1.5\,\rm dex$ for the cluster. Similar feature also exists in the metal-poor globular clusters NGC~6397 and NGC~6752 \citep{Moehler2025A&A...693A.136M}. A blue horizontal-branch exists in the metal-rich ($\rm [M/H]=-0.48\,\rm dex$) globular cluster NGC~6388, and \citet{Moehler2025A&A...693A.136M} explained this feature as the Helium enrichment.

There are $\sim$2000 BHB stars have $\rm [Fe/H]>-1\,\rm dex$ in our sample, and they appear as a bump around $\rm [Fe/H]=-0.5\,\rm dex$ in the distribution of [Fe/H] (see the right panel of Figure~\ref{fig:HIST_params}). These metal-rich BHB stars ($\rm [Fe/H]>-1$) fall on the boundary between BHB and BS stars. Most of these metal-rich stars display discrepant types among various Balmer lines, and some of them could be contaminant BS stars. These stars lie closer to the BHB stars than to the BS stars in the Kiel diagram, and for the sake of completeness, they are retained in our sample. In the catalogs of \citet{Brown2008AJ....135..564B} and \citet{Ju2025ApJS..276...12J}, there are also BHB stars with $\rm [Fe/H]>-1\,\rm dex$, and \citet{Guo2025A&A...702A..11G} noted that nearly all of the BHB stars with $\rm [Fe/H]>-1\,\rm dex$ from \citet{Ju2025ApJS..276...12J} have kinematics consistent with being disk members.

\begin{figure*}[!t]
    \centering
    \includegraphics[width=\textwidth]{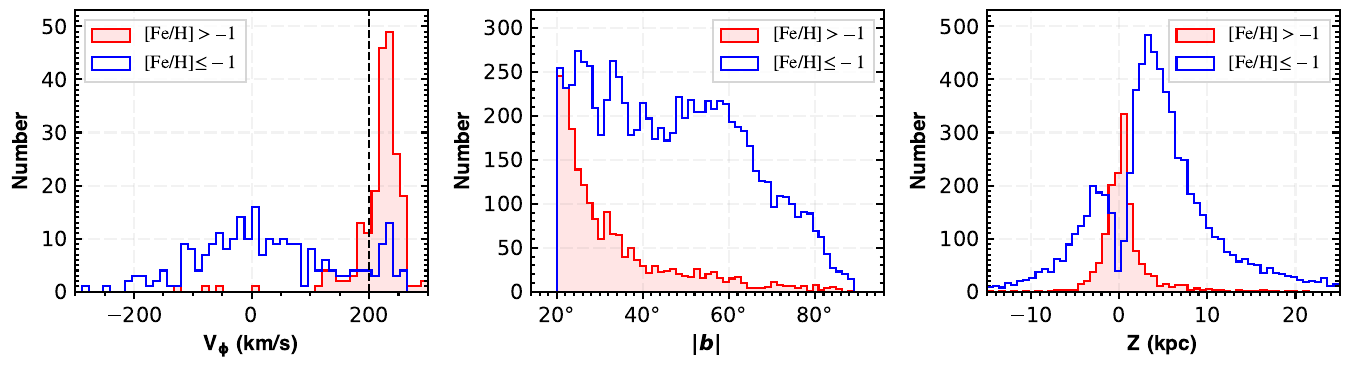}
    \caption{Histograms showing the distributions of $V_{\phi}$ (left panel), $|b|$ (middle panel), and $Z$ (right panel) for BHB stars with $\rm [Fe/H]>-1\,dex$ (red filled histogram) and $\rm [Fe/H]\leq-1\,dex$ (blue open histogram). The vertical dashed line in the left panel corresponds to $V_{\phi}=200\,\rm km\,s^{-1}$, and is used to separate the sample into disk ($V_{\phi}\ge200\,\rm km\,s^{-1}$) and halo ($V_{\phi}<200\,\rm km\,s^{-1}$) populations.}
    \label{fig:hist_vphi}
\end{figure*}

Cross-matching our BHB sample with the radial velocity (RV) catalog from \citet{Zhang2021ApJS..256...14Z}, and combining with the astronomical measurements from Gaia DR3 \citep{Gaia-Collaboration2023A&A...674A...1G}, we calculate azimuthal velocities ($V_{\phi}$) for 463 BHB stars, including 208 stars have $\rm [Fe/H]>-1\,\rm dex$ and 255 stars have $\rm [Fe/H]\leq-1\,\rm dex$. $V_{\phi}$ values for the two samples are compared in the left panel of Figure~\ref{fig:hist_vphi}. Of the 208 metal-rich BHB stars, 169 ($>$80\%) have $V_{\phi}>200\,\rm km\,s^{-1}$, and thus are likely to be disk members\footnote{Higher $V_{\phi}$ values generally indicate significant rotational motion, characteristic of disk stars, while lower $V_{\phi}$ values are consistent with halo stars \citep{Tian2019ApJ...871..184T}. \citet{Guo2025A&A...702A..11G} adopted metallicity dependent $V_{\phi}$ values to separate disk members and halo members. In this work, we adopt a unique value of $200\,\rm km\,s^{-1}$ to separate disk and halo populations for simplicity.}, while only 33 ($<$15\%) of the metal-poor stars belong to the disk population. In the middle and right panels of Figure~\ref{fig:hist_vphi}, we also display the distributions of $|b|$ and $Z$ for the two populations, respectively. As shown, most of the metal-rich stars locate at lower latitudes, and are closer to the Galactic plane, than the metal-poor stars. Both spatial distributions and kinematics indicate the metal-rich BHB stars are likely to belong to the disk population, similar as in \citet{Guo2025A&A...702A..11G}.

\section{Summary}\label{sec:sum}

Large samples of BHB stars with determined atmospheric parameters are essential for studying the kinematics and substructures of the Galaxy, as well as investigating stellar evolution. In this work, we performed a systematic search for BHB stars via the $D_{0.2}$ method and the scale width-shape method, based on LAMOST DR11. The atmospheric parameters are determined for the BHB sample. The main contributions are summarized as follows.

\begin{enumerate}
\item By analyzing the three Balmer lines, and using the $D_{0.2}$ method and the scale width-shape method, we identify 13\,988 BHB spectra, corresponding to 10\,236 unique BHB stars, based on LAMOST DR11.
\item Atmospheric parameters are determined for the BHB stars, using the machine learning method \texttt{SLAM}. We demonstrate that including color indices in the spectral labeling can effectively break the degeneracy between effective temperature and surface gravity, and provide more reliable estimate of the stellar labels. The improvement is more significant for the lower-SNRg spectra.
\item By comparing with previous studies \citep{Xue2011ApJ...738...79X,Vickers2021ApJ...912...32V,Ju2024ApJS..270...11J,Culpan2024A&A...685A.134C}, we estimate an identification rate of $\sim$80\%\textendash90\% of our sample. Through comparisons in GCs, we estimate a contamination rate of $\lesssim$10\% for our sample, and this is confirmed by placements of the BHB stars in the CaMD.
\item We note a bump in the distribution of [Fe/H], and by analyzing the spatial distributions and kinematics of these metal-rich BHB stars, we demonstrate that most of them belong to the disk population.
\item As a by-product, we also provide a list of 4282 BS stars with determined atmospheric parameters. These BS stars serve as a good negative sample to remove contaminants from BHB samples.
\end{enumerate}

\begin{acknowledgments}
This study is supported by the National Natural Science Foundation of China under grant No. 12573026, the China Manned Space Program with grant No. CMS-CSST-2025-A13, the Hebei Natural Science Foundation under grant No. A2023205036 and A2026205025. X.-L.W. acknowledges the support by the Science Foundation of Hebei Normal University (grant No. L2024B56) and S\&T Program of Hebei (grant No. 22567617H). X.-L.W. is supported by the LAMOST FELLOWSHIP as a Youth Researcher, which is supported by the LAMOST Operation and Development Center, National Astronomical Observatories, Chinese Academy of Sciences (NAOC). This work has made use of data from the European Space Agency (ESA) mission {\it Gaia} (\url{https://www.cosmos.esa.int/gaia}), processed by the Gaia Data Processing and Analysis Consortium (DPAC, \url{https://www.cosmos.esa.int/web/gaia/dpac/consortium}). Funding for the DPAC has been provided by national institutions, in particular the institutions participating in the Gaia Multilateral Agreement. This work made use of the data from LAMOST (Large Sky Area Multi-Object Fiber Spectroscopic Telescope, also known as the Guoshoujing Telescope) (\url{https://cstr.cn/31118.02.LAMOST}). LAMOST is a Chinese national mega-science facility, operated by National Astronomical Observatories, Chinese Academy of Sciences.
\end{acknowledgments}


\bibliography{References}{}
\bibliographystyle{aasjournal}

\end{document}